\documentclass[a4paper,fleqn,usenatbib]{mnras}

\usepackage{newtxtext,newtxmath}
\usepackage{newtxtext,newtxmath}
\usepackage[T1]{fontenc}
\usepackage{ae,aecompl}

\usepackage{graphicx}	% Including figure files
\usepackage{amsmath}	% Advanced maths commands
\usepackage{amssymb}	% Extra maths symbols

\usepackage{color}
\usepackage{epsfig}
\usepackage{comment}

\allowdisplaybreaks[1]

\usepackage{here}

\newcommand{\msun}{{\rm M}_\odot}

\newcommand{\zsun}{Z_\odot}

\newcommand{\cc}{{\rm cm}^{-3}}

\newcommand{\msunyr}{{\rm M}_\odot~{\rm yr}^{-1}}

\newcommand{\K}{{\rm K}}
\newcommand{\beq}{\begin{equation}}
\newcommand{\eeq}{\end{equation}}

\title[Rapid BH growth and powerful outflows]
{Hyper-Eddington accretion flows onto black holes accompanied by powerful outflows}

\author[E. Takeo, K. Inayoshi, and S. Mineshige]{
Eishun Takeo$^{1}$\thanks{takeo@kusastro.kyoto-u.ac.jp},
Kohei Inayoshi$^{2,3}$\thanks{inayoshi@pku.edu.cn, corresponding author}, and
Shin Mineshige$^{1}$\\
$^{1}$Department of Astronomy, Graduate School of Science, Kyoto University, 
Kitashirakawa, Oiwakecho, Sakyo-ku, Kyoto, 606-8502, Japan\\
$^{2}$The Kavli Institute for Astronomy and Astrophysics, Peking University, 
5 Yiheyuan Road, Haidian District, Beijing 100871, P. R. China \\
$^{3}$Department of Astronomy, School of Physics, Peking University, Beijing 100871, P. R. China\\
}

\date{Accepted XXX. Received YYY; in original form ZZZ}

\pubyear{2020}

\begin{document}
\label{firstpage}
\pagerange{\pageref{firstpage}--\pageref{lastpage}}
\maketitle

\begin{abstract}
We perform two-dimensional radiation hydrodynamical simulations of accretion flows onto black holes (BHs)
at the nuclei of protogalaxies, and study the impact of mechanical and radiative feedback on rapid growth of BHs.
The outflows deposit mass, momentum and energy into the surrounding medium and prevent mass accretion 
onto the BH, resulting in the reduction of radiative output.
We find that when the BH is embedded in a dense gas core, ionizing radiation attenuated by inefficient BH feeding 
owing to mechanical feedback
hardly affects the gas dynamics at the BH gravitational sphere of influence, from which intense inflows of neutral gas 
occur at rates substantially exceeding the Eddington limit without impeded by photoionization and heating.
Since mechanical power of outflows driven by the rapidly accreting BH is sufficiently strong, 
bipolar outflows completely evacuate the surrounding gas in the polar region but mass inflows through the equatorial region
maintain the BH accretion rate as high as $\sim 300-10^3~\dot{M}_{\rm Edd}$, which is reduced by one order of magnitude from 
those with radiative feedback alone.
Furthermore, we find that the critical gas density required for rapid accretion is lower by a factor of $\sim 3$
nearly independently of BH mass, when mechanical feedback is considered.
By studying the dependence on outflow model parameters (e.g., opening angle, mass-loading degree into outflows, velocity), 
we conclude that contrary to naive expectation, the existence of stronger outflow leads to the transition to rapid accretion phases 
more efficiently.
Rapidly growing BHs inject mechanical power with $\sim 0.1-1\%$ of the radiative luminosity into their host galaxy scales, 
which is used for cosmological simulations.
\end{abstract}


%%%%%%%%%%%%
%	1. Introduction	 %
%%%%%%%%%%%%
\section{Introduction}

Supermassive black holes (SMBHs) with masses of $\gtrsim 10^9~\msun$ 
\citep[e.g.][]{Fan_2004,Mortlock_2011,Wu_2015,Banados_2018,Matsuoka+2018a, Matsuoka+2018b,
Matsuoka+2018c,Onoue+2019} as the central engine of bright quasars at high redshifts $z \gtrsim 6$ 
(or less than $\sim 1~{\rm Gyr}$ after the Big Bang) require their rapid growth in the early Universe.
Although SMBHs play crucial roles on the history of their host galaxies
\citep[e.g.][reference theirin]{Silk_Rees_1998,King_2003,Murray_2005, Kormendy_Ho_2013}, 
their formation and growth mechanisms are still unclear.

To assemble such monster SMBHs within a certain short timescale, 
several seeding models have been proposed by many authors 
\cite[e.g.,][references therein]{Volonteri2012,Haiman2013,Inayoshi+2019ARAA}.
One possible candidate of the seed is the stellar-mass BH with $\sim 10-100~\msun$ left behind by collapse of 
the first generation of stars.
Assuming that those remnant BHs grow via gas accretion at the corresponding Eddington rate with
a 10\% of the radiative efficiency, the timescale required to form $\gtrsim 10^9~\msun$ SMBHs becomes 
as long as $\sim 1$ Gyr, which is comparable to the age of the high-redshift Universe.
Therefore, this naive idea could work only if rapid BH accretion with a high duty cycle 
was sustained over $7-8$ orders of magnitude growth in mass \citep[][]{TH09}.

An alternative candidate is more massive seed BHs with $\sim 10^4-10^6~\msun$ formed by 
direct collapse of supermassive stars or runaway stellar mergers in a dense metal-poor star cluster
\citep{bromm_loeb_2003,DevecchiVolonteri2009,shang_2010,Rega+2014,IOT14,Latif_2016,
Tagawa+2019,ChonOmukai2020}.
Such massive seeds are expected to form in high-$z$ protogalaxies under peculiar conditions
such as the existence of strong H$_2$ photodissociating Lyman-Werner radiation 
\citep{Dijkstra_2008,Johnson_2013,sugimura_2014,Visbal_2014,WHB_2017},
high baryon-dark matter streaming velocity, rapid mergers of dark-matter halos, or 
some combination of these effects \citep{TanakaLi2014,inayoshi_2018,Wise+2019}. 
Even giving the massive seeds a head start towards SMBHs at $z\gtrsim 6$, the growth timescale would be reduced 
only by a factor of two under the assumption of Eddington-limited accretion.
Therefore, it is worth exploring the possibility that seed BHs embedded in protogalaxies can grow faster 
breaking the Eddington limit (though the duration should be short enough to be consistent with the Soltan argument) 
and addressing whether the conditions to trigger super-Eddington accretion are consistent with those 
of their seeding models.

Rapid gas accretion onto compact objects has been investigated by numerous studies. 
There are several lines of observational evidence to show both stellar-mass BHs and SMBHs
can be super-Eddington accretors; e.g., some ultra-luminous X-ray sources \citep[e.g.,][]{King+2001,Watarai_2001}
and narrow-line Seyfert-1 galaxies \citep[e.g.,][]{Wang+1999, Mineshige+2000}.
In theory, the properties of super-Eddington accretion have been explored by analytical and numerical work.
In the recent decade, radiation (magneto-)hydrodynamical simulations have shown that super-Eddington 
accreting flows can feed the central BH as long as a sufficient mount of gas already exists or is efficiently supplied 
to the vicinity of the BH event horizon scale of $\sim 10^{2-3}~R_{\rm Sch}$ 
\citep[see also][]{ohsuga+2009,ohsuga_mineshige2011,Jiang_2014,McKinney_2014,Sadowski_2015,takahashi2016,Jiang_2019},
where $R_{\rm Sch} \equiv 2GM_{\rm BH}/c^2$ is the Schwarzschild radius, 
$M_{\rm BH}$ is the BH mass, and $c$ is the speed of light.
However, accreting matter at super-Eddington rates release a huge amount of gravitational potential energy 
as radiation and/or in outflows.
Emergent ionizing photons are likely to propagate outward and heat the inflowing gas from 
the BH gravitational sphere of influence (hereafter, the Bondi radius $R_{\rm B}$).
Since gas-pressure gradient force (and partially radiation force exerted through electron scattering) 
in the ionized region overcomes gravity of the central BH, gas accretion from larger radii to the nuclear 
BH is strongly suppressed.
This quiescent phase lasts until the size of the ionized bubble substantially shrinks because of 
the combination of radiative recombination and lower radiative luminosity from the BH.
As a result, radiative feedback due to photoionization and heating limits the BH feeding rate significantly 
below the Eddington accretion rate $\dot{M}_{\rm Edd} ~[\equiv L_{\rm Edd}/(0.1c^2)]$
\citep[e.g.,][]{Ciotti_Ostriker_2001,Milosavljevic_2009a,Milosavljevic_2009b,Alvarez_2009,
Park_Ricotti_2011,Park_Ricotti_2012,Jeon+2012},
where $L_{\rm Edd}$ is the Eddington luminosity.

However, when the BH is embedded in a dense gas cloud so that the gas inflowing rate from $R_{\rm B}$ 
exceeds $\sim 500~\dot{M}_{\rm Edd}$, the inflowing gas structure approaches a steady state without 
time-dependent oscillations, yielding hyper-Eddington accretion \citep{inayoshi+16}. 
The critical accretion rate required to realize hyper-Eddington accretion is rewritten as 
\begin{equation}
n_{\infty} \gtrsim 10^4~\cc ~\left(\frac{M_{\rm BH}}{10^5~\msun}\right) \left(\frac{T_\infty}{10^4~\K}\right)^{3/2},
\label{eq:super_Edd}
\end{equation}
where $n_\infty$ and $T_\infty$ are density and temperature of the surrounding gas.
Note that this inequality is satisfied when the size of an ionized region surrounding an accreting BH, 
$R_{\rm ion}$ (well approximated by the Str\"{o}mgren radius) is smaller than the Bondi radius.
In this case, since radiation emitted from the BH hardly ionizes and heats the ambient gas,
gas inflows that begins to occur from $\sim R_{\rm B}$ are not prevented by radiative feedback, 
but rather lead to collapse of the ionized region.
Several follow-up radiation hydrodynamical (RHD) simulations have shown that when the radiation is 
emitted preferentially towards the polar regions, rapid gas accretion is allowed through the equatorial region 
with strong inward ram pressure that leads to collapse of the bipolar ionized region (\citealt{takeo_2018};
see also \citealt{sugimura_2017}).
We also note that the size of an ionized region depends on spectra of radiation.
\cite{takeo_2019} re-evaluated the transition criteria, taking into account more realistic radiation spectra 
associated with the properties of nuclear accretion discs that depend on the BH mass and accretion rate.
They found that the required density is lowered for less massive BHs because the mean photon energy 
becomes too high to ionize and heat the ambient gas.
In addition, when the accreting gas is slightly polluted by heavy elements ($\lesssim 0.01~\zsun$) and 
contains dust grain, the spectral shape of emergent radiation is softened due to attenuation of ultraviolet photons, making 
the ionized regions smaller \citep{Yajima+2017, Toyouchi+2019}.

The next question is whether mechanical feedback associated with outflows\footnote{
In this paper, ``outflow'' is defined as mass flow at a non-relativistic velocity.
Relativistic (and highly collimated) outflows are phrased as jets.} 
launched from an accretion disc affects the properties of gas inflows from outside $R_{\rm B}$.
So far, various mechanisms driving outflows have been proposed.
In fact, most numerical simulations of super-Eddington accretion flows show the presence of 
significant mass outflows driven by radiation pressure 
\citep[e.g.,][]{ohsuga+05,kawashima+2009,Jiang_2014,Sadowski_2015,Jiang_2019}.
Even with a sub-Eddington accretion rate, strong outflows can be generated owing to the large-scale 
poloidal magnetic field \citep[e.g.,][]{Blandford_Payne_1982,Li_2019}, large line opacity 
\citep[e.g.,][]{Proga_2000,Nomura_2016,Nomura_2018}, and by extracting the BH spin \citep{Blandford_1977}.
Applying those results to cosmological simulations of galaxy evolution,
numerous authors have been investigating 
the impact of mechanical feedback on BH accretion and star formation in galactic scales
at relatively low redshifts \citep[e.g.,][]{Dubois+2010,ostriker+2010,Novak_2011,Choi+2012,Yuan+2018,Qiu+2019}.

Recently, \cite{Regan+2019} have studied the effect of bipolar jets launched from an accreting seed BH 
onto gas inflows in an atomic cooling halo, performing cosmological simulations that resolve the BH 
gravitational influence radius. 
They found that the jets evacuate the central $\sim 0.1~{\rm pc}$ around the BH $(\simeq 0.1~R_{\rm B})$ 
and limit the accretion rate below the Eddington value.
In fact, since the jet injection scale is not spatially resolved well, the jets are not sufficiently collimated at the bottom
but prevent gas inflows even through the equatorial plane (i.e., perpendicular to the gas 
angular momentum vector).
This leads to overestimate of the mechanical feedback effect.
On the other hand, their simulations do not take into account radiative feedback (photoionization and heating),
which evacuates gas around the BH and could assist the outflow component in expanding 
outward and affecting the gas inflow from larger radii.
In this sense, the impact of mechanical feedback on gas inflows is underestimated in their simulations 
(see discussion in \S\ref{sec:out_para}).

In this paper, we explore the conditions required for hyper-Eddington accretion onto a BH 
when both radiative and mechanical feedback operate simultaneously, performing two-dimensional 
hydrodynamical simulations with multi-frequency radiation transfer.
We conduct a comprehensive survey on the parameter dependence of outflow models, varying the outflow 
opening angle, mass-loading degree into outflows, velocity of outflows, and density of gas surrounding the BH.
To model mechanical feedback, we adopt a phenomenological model proposed by \cite{ostriker+2010},
while radiative feedback is treated by adopting the standard and slim disc model 
\citep{SS_1973,abramowicz_1988,watarai_2006} as in \cite{takeo_2019}.
We find that the flow structure consists of two distinct parts in the early sub-Eddington phase; 
the bipolar outflowing region heated up to $T\sim 10^{6-7}~\K$ due to strong shock
and the equatorial inflowing region where ionized gas is mildly heated to $T\sim 10^5~\K$
due to photoionization.
When the ambient gas density exceeds a critical threshold, as in the cases where only 
radiative feedback is included \citep{inayoshi+16,takeo_2018}, the mass accretion rate onto the nuclear
region rises to a hyper-Eddington value.
Since mechanical power of outflows driven by the rapidly accreting BH is sufficiently strong, 
bipolar outflows completely evacuate the surrounding gas in the polar region and reduce 
the mass inflow (BH accretion) rate by a factor of $\approx 3-13$ ($\approx 6-26$, respectively)
from the case without mechanical feedback.
Furthermore, we find that the critical gas density required for hyper-Eddington accretion is 
reduced by a factor of $\sim 3$ and the transition occurs in a shorter dynamical timescale
when mechanical feedback is modelled in the simulations.
In fact, the effects that alleviate the transition to rapid accretion tend to be more prominent
as the outflow is stronger, i.e., a wider opening angle, higher mass-loading factor, and higher outflow velocity.
This is because suppression of BH accretion owing to outflows reduces the radiative output 
from the nuclear BH, leading to hyper-Eddington accretion.

% KI: should be revised
The rest of this paper is organized as follows.
In \S\ref{sec:2}, we describe our numerical setup and list the models we study.
In \S\ref{sec:3}, we present the simulation results and discuss their dependence on the outflow properties.
In \S\ref{sec:4}, we derive the conditions required for the transition to hyper-Eddington accretion
and briefly discuss Impact of BH feedback on the host galaxy evolution.
Finally, we summarize our findings in \S\ref{sec:5}.

%%%%%%%%%%
%	2. Method	    %
%%%%%%%%%%
\section{Methods}
\label{sec:2}

In this paper, we study the properties of rapid gas inflows onto a BH supplied from larger scales ($\gg R_{\rm Sch}$),
performing axisymmetric two-dimensional radiation hydrodynamical simulations.
To investigate the effects of radiative and mechanical feedback on the gas dynamics,
our simulations resolve the BH gravitational sphere of influence, the so-called Bondi radius \citep{Bondi_1952}, defined by
\begin{equation}
R_{\rm B}\equiv \frac{GM_{\rm BH}}{c_{\infty}^2} \simeq 1.97\times10^{19} \ {\rm cm} \
\left(\frac{M_{\rm BH}}{10^5~\msun}\right)
\left(\frac{T_{\infty}}{10^4~\K}\right)^{-1},
\label{eq:bondi_radius}
\end{equation}
where $c_{\infty} \equiv \sqrt{\gamma \mathcal{R} T_{\infty} /\bar{\mu}}$ is the sound speed,
$\gamma$ is the specific heat ratio, 
$\mathcal{R}$ is the gas constant, and
$\bar{\mu}$ is the mean molecular weight.
Within the Bondi radius, the BH gravitational energy dominates over the thermal energy of the gas,
and thus gas accretion begins to occur unless feedback associated with BH feeding plays an important role.
We also define the Bondi accretion rate for isothermal gas as 
\begin{align}
  \dot{M}_{\rm B} &\equiv \pi e^{3/2} \rho_\infty \frac{G^2 M_{\rm BH}^2}{c_{\infty}^3},\\
  & \simeq  7\times 10^3~\dot{M}_{\rm Edd} \left( \dfrac{M_{\rm BH}}{10^5~\msun} \right) \left( \dfrac{n_{\infty}}{10^5~\cc} \right)
  \left( \dfrac{T_{\infty}}{10^4~\K} \right)^{-3/2}.
  \label{bondi_rate}
\end{align}
Throughout this paper, the Bondi radius and rate are calculated by setting 
$\gamma=1$, $\bar{\mu}=1.23$ and $T_{\infty}=10^4 {\rm K}$ as reference values,
although those values are simulated self-consistently.

%%%
\subsection{Basic equations}
\label{sec:BE}

We solve the axisymmetric two-dimensional hydrodynamical equations
using a code developed in \cite{takahashi_ohsuga_2013} which compute the flux with 
the Harten-Lax-vanLeer Riemann solver \citep{harten_1983} and 
ensures the second order accuracy in space and time \citep{van_leer_1977}.
Here we employ spherical coordinates of $(r, ~\theta, ~\phi)$ 
with the polar axis ($\theta = 0$ and $\pi$) perpendicular to the equatorial plane ($\theta =\pi/2$).

The hydrodynamical equations are the following:
the equation of continuity
\begin{equation}
  \frac{\partial \rho}{\partial t} + \nabla\cdot(\rho {\boldsymbol v}) = \dot{\rho}_{\rm out},
  \label{renzoku}
\end{equation}
the equations of motion
\begin{equation}
  \frac{\partial \left(\rho v_r\right)}{\partial t} + \nabla\cdot(\rho v_r {\boldsymbol v}) 
  = -\frac{\partial p}{\partial r} + \rho\left( \frac{v_{\theta}^2}{r} +\frac{v_{\phi}^2}{r} \right) - \rho \frac{\partial \psi}{\partial r}+ f_{\rm rad} + \dot{p}_{\rm out} ,
  \label{eomr}
\end{equation}
\begin{equation}
  \frac{\partial \left(\rho rv_{\theta}\right)}{\partial t} + \nabla\cdot(\rho r v_{\theta} {\boldsymbol v}) 
  = -\frac{\partial p}{\partial \theta} + \rho v_{\phi}^2\cot{\theta},
  \label{eomt}
\end{equation}
\begin{equation}
  \frac{\partial \left(\rho rv_{\phi}\sin{\theta}\right)}{\partial t} + \nabla\cdot\left(\rho r v_{\phi}\sin{\theta} {\boldsymbol v}\right) = 0,
  \label{eomp}
\end{equation}
and the energy equation
\begin{equation}
  \frac{\partial e}{\partial t} + \nabla\cdot[(e+p){\boldsymbol v}] = -\frac{GM_{\rm BH}\rho}{r^2}v_r -\Lambda + \Gamma + \dot{e}_{\rm out} .
  \label{eneeq}
\end{equation}
where $\rho$ is the gas density, ${\boldsymbol v}=(v_r,v_{\theta}, v_{\phi})$ is the velocity vector, 
$p$ is the gas pressure, and $f_{\rm rad}$ is the outward radiation force onto the radial direction.
We take into account gravitational force of the central BH ($r=0$) and neglect the gas self-gravity. 
Since the general relativistic effect is negligible around $R_{\rm B}(\gg R_{\rm Sch})$, the gravitational potential is given by 
$\psi=-GM_{\rm BH}/r$.
The total energy per volume (in erg cm$^{-3}$) is defined by 
$e \equiv e_{\rm int}+\rho |{\boldsymbol v}|^2/2$, 
$e_{\rm int}$ is the gas internal energy density, 
$\Lambda$ is the radiative cooling rate, 
and $\Gamma$ is the radiative heating rate.
We assume the equation of state of ideal gas as $p=(\gamma-1)e_{\rm int}$ for $\gamma=5/3$. 
To ensure the mass, momentum and energy conservation, we add source terms associated with 
mechanical feedback ($\dot{\rho}_{\rm out}$, $\dot{p}_{\rm out}$, and $\dot{e}_{\rm out}$).

We consider cooling processes associated with ${\rm H, He, He^+}$ atoms and 
free-free emission \citep{glover+07}, adopting the optically-thin cooling rates.
We solve chemical reaction networks including six species of 
$\rm H, H^+, He, He^+, He^{++}$, and ${\rm e^-}$. 
The abundance of He nuclei relative to H nuclei is set to $8.33\times10^{-2}$.
Here we adopt the chemical processes of photoionization, collisional ionization and 
radiative recombination \citep{abel+97,glover+07}, including the secondary ionization (see Section \ref{sec:RT}). 
Since photoionization due to diffusive recombination photons is negligible,
we adopt the on-the-spot approximation 
where the case A recombination rate is replaced by the case B rate. 
In order to solve the basic equations stably, the source terms associated with radiative cooling/heating
in the energy equation and the chemical reaction networks are updated with a fully implicit method.
Throughout our simulations, we set the time step to the Courant timescale 
(the Courant number is set to 0.4).

%%%
\subsection{Mechanical feedback}

To study the impact of mechanical feedback due to outflows on gas inflow, we adopt a phenomenological 
model proposed by \citep{ostriker+2010}, where conservation of mass, momentum and energy is taken into account.
Defining the mass inflow through the innermost radius of $r_{\rm min}$ (see Section \ref{sec:IBC}) as
\begin{equation}
\dot{M}_{\rm in} = \int_{v_r < 0}{\rm d}\Omega~ r_{\rm min}^2~\rho |v_r|,
\end{equation}
and the mass loading factor into outflows as 
$\beta_{\rm out} \equiv \dot{M}_{\rm out} / \dot{M}_{\rm BH}$,
we obtain the mass outflow rate and BH accretion rate from mass conservation,
\begin{equation}
\dot{M}_{\rm out} = \dfrac{\beta_{\rm out}}{1+\beta_{\rm out}} \dot{M}_{\rm in},
\end{equation}
\begin{equation}
\dot{M}_{\rm BH} = \dfrac{1}{1+\beta_{\rm out}} \dot{M}_{\rm in}.
\label{BH_acc_rate}
\end{equation}
Here, it is worthy noting the following two things that are natural consequences from 
mass conservation but sometimes misinterpreted.
Note that mass conservation ensures that the outflow mass rate never exceeds the inflow rate
even in the limit of $\beta_{\rm out} \gg 1$.
In the same limit, the BH accretion rate goes to zero no matter how much mass inflow rate exists 
at $r=r_{\rm min}$.

Next, we define the total momentum and energy input due to outflows as
\begin{equation}
\dot{P}_{\rm out} = \dot{M}_{\rm out} v_{\rm out},
\end{equation}
\begin{equation}
\dot{E}_{\rm out} =\dfrac{1}{2} \dot{M}_{\rm out}v_{\rm out}^2, 
\end{equation}
where $v_{\rm out}$ is the outflow velocity injected at $r=r_{\rm min}$.
Throughout this paper, we treat $\beta_{\rm out}$ and $v_{\rm out}$ as free parameters
varying those values and discuss the dependence on the result (see \S\ref{sec:out_para}).
The mass, momentum, and energy of outflows are injected at the inner-most grid 
as source terms $\dot{\rho}_{\rm out}$, $\dot{p}_{\rm out}$ and $\dot{e}_{\rm out}$ in an anisotropic way:
\begin{eqnarray}
\dot{\rho}_{\rm out} = \dfrac{\dot{M}_{\rm out}}{4\pi r_{\rm min}^2 \mathcal{C}} \mathcal{G}(\theta) \delta(r-r_{\rm min}), \\
\dot{p}_{\rm out} = \dfrac{\dot{P}_{\rm out}}{4\pi r_{\rm min}^2 \mathcal{C}} \mathcal{G}(\theta) \delta(r-r_{\rm min}), \\
\dot{e}_{\rm out} = \dfrac{\dot{E}_{\rm out}}{4\pi r_{\rm min}^2 \mathcal{C}} \mathcal{G}(\theta) \delta(r-r_{\rm min}).
\end{eqnarray}
Here, $\mathcal{G}(\theta)$ characterizes the angular dependence of outflows as
\begin{equation}
\mathcal{G}(\theta) =
\begin{cases}
1 &, ~0^{\circ} \leq \theta \leq \widetilde{\theta}_{\rm out} \\
{\rm exp}\left(-\left(\dfrac{\theta - \widetilde{\theta}_{\rm out}}{\Delta\theta} \right)^2 \right) &, 
~\widetilde{\theta}_{\rm out}\leq \theta \leq 90^{\circ}, 
\end{cases}
\label{angular_outflow}
\end{equation}
where $\mathcal{C} \equiv \int_{0}^{\pi/2}{\rm d}\theta\sin{\theta}~\mathcal{G}(\theta)$, 
$\delta$ is the Dirac delta-function, $ \widetilde{\theta}_{\rm out} = \theta_{\rm out}-2\Delta\theta$, 
and we set $\Delta\theta = 6^{\circ}$ \citep[see also][]{sugimura_2017,ciotti+2017}.

%%%%%%%%%
%	Table 1	%
%%%%%%%%%
\begin{table*}
\begin{center}
\caption{
Simulation Runs and Input Parameters. 
Column (1) model ID, (2) BH mass, (3) ambient gas density, (4) outflow opening angle, (5) mass loading factor,
(6) outflow velocity, and (7) the transition epoch in units of the dynamical time at the Bondi radius; 
$t_{\rm dyn}=8.4\times 10^5~(M_{\rm BH}/10^5~\msun)$ yr.
For all the models, isotropic radiation is assumed in the early stage before the transition to rapid accretion occurs.
For the first three models (A45M5-), we extend the simulations to the late stage after the transition epoch, 
varying the feedback model parameters: 
anisotropic radiation with an opening angle of $\theta_{\rm rad}=45^\circ$ (A45M5), 
anisotropic radiation with $\theta_{\rm rad}=45^\circ$ and $\beta_{\rm out}=10$ (A45M5-B10), and
isotropic radiation but assuming a diluted blackbody spectrum (A45M5-B1-iso).
In Model RAD, only radiative feedback is considered as reference.}
\begin{tabular}{lccccccc}
\\
Model & $M_{\rm BH}~(\msun)$ &  $n_{\infty}~(\cc)$ & $\theta_{\rm out}~(^{\circ})$ &$\beta_{\rm out}$ & 
$v_{\rm out}~({\rm km~s^{-1}})$ & $t_{\rm tran}~(t_{\rm dyn})$ & Reference \\
\hline 
A45M5 (fiducial) & $10^5$ & $1\times10^5$ & $45$ & $1$ & $1000$ & $0.96$ & \S\ref{sec:angle}\\
A45M5-B10 & $10^5$ & $1\times10^5$ & $45$ & $10$ & $1000$ & $-$ & ~~~\S\ref{sec:highbeta} \\
A45M5-B1-iso & $10^5$ & $1\times10^5$ & $45$ & $1$ & $1000$ & $-$  & ~~~\S\ref{sec:iso_rad}\\
\hline
A60M5 & $10^5$ & $1\times10^5$ & $60$ & $1$ & $1000$ & $0.84$ & \S\ref{sec:out_para} \\
A80M5 & $10^5$ & $1\times10^5$ & $80$ & $1$ & $1000$ & $0.62$ & \S\ref{sec:out_para}  \\
V500M5 & $10^5$ & $1\times10^5$ & $80$ & $1$ & $500$ & $0.71$ & \S\ref{sec:out_para} \\
V333M5 & $10^5$ & $1\times10^5$ & $80$ & $1$ & $333$ & $0.75$ & \S\ref{sec:out_para} \\
V200M5 & $10^5$ & $1\times10^5$ & $80$ & $1$ & $200$ & $0.88$ & \S\ref{sec:out_para} \\
B01M5 & $10^5$ & $1\times10^5$ & $80$ & $0.1$ & $1000$ & $0.79$ & \S\ref{sec:out_para} \\
B001M5 & $10^5$ & $1\times10^5$ & $80$ & $0.01$ & $1000$ & $1.3$ & \S\ref{sec:out_para} \\
\hline
T34M5 & $10^5$ & $3\times10^4$ & $80$ & $1$ & $1000$ & $1.7$ & \S\ref{sec:tran}\\
T24M5 & $10^5$ & $2\times10^4$ & $80$ & $1$ & $1000$ & $2.5$  & \S\ref{sec:tran}\\
T14M5 & $10^5$ & $1\times10^4$ & $80$ & $1$ & $1000$ & $-$ & \S\ref{sec:tran} \\
T33M5 & $10^5$ & $3\times10^3$ & $80$ & $1$ & $1000$ & $-$ & \S\ref{sec:tran} \\
T18M1 & $10$ & $1\times10^8$ & $80$ & $1$ & $1000$ & $0.79$ & \S\ref{sec:tran} \\
T57M1 & $10$ & $5\times10^7$ & $80$ & $1$ & $1000$ & $1.3$ & \S\ref{sec:tran} \\
T37M1 & $10$ & $3\times10^7$ & $80$ & $1$ & $1000$ & $1.8$ & \S\ref{sec:tran} \\
T17M1 & $10$ & $1\times10^7$ & $80$ & $1$ & $1000$ & $-$ & \S\ref{sec:tran} \\
\hline
RAD (w/o outflows)& $10^5$ & $1\times10^5$ & $-$ & $-$ & $-$ & $1.3$ \\
\hline
\end{tabular}
\label{models}
\end{center}
\end{table*}

%%%
\subsection{Radiation transfer and radiative feedback}
\label{sec:RT}

To estimate the radiative heating rate $\Gamma$, ionization rate $k_{\rm ph}$, 
and radiation force $f_{\rm rad}$, we need to solve for radiation quantities in the accreting gas. 
In this paper, we adopt the same treatment of radiation transfer as described in
our previous work \citep{inayoshi+16, takeo_2018, takeo_2019}.

The multi-frequency radiative transfer equation is give by
\begin{equation}
\frac{1}{r^2}\frac{{\rm d}}{{\rm d}r}(r^2F_{\nu}) = -\rho\kappa_{\nu}cE_{\nu},
\label{rteq}
\end{equation}
where $F_{\nu}$ is the radiation flux, $E_{\nu}$ is the radiation energy density, and 
$\kappa_{\nu}$ is the absorption opacity.
Here, the radiation field is assumed to be steady because the light crossing time is much shorter 
than the hydrodynamical time step.
We take into account only the radial component of radiation flux because non-radial components 
are negligible (see \S4 in \citealt{takeo_2018} for more details).
The frequency range is set to $h\nu_{\rm min}(=13.6~{\rm eV}) \leq h\nu \leq h\nu_{\rm max}(=100~{\rm keV})$,
where $h$ is the Planck constant.
Inside the ionized region, the gas is optically thin even to ionizing photons and thus $E_{\nu} \simeq F_{\nu}/c$.
Therefore, the radiation transfer equation is approximated in a simple form:
\begin{equation}
\frac{1}{r^2}\frac{{\rm d}}{{\rm d}r}(r^2F_{\nu}) = -\rho\kappa_{\nu}F_{\nu}.
\label{rteq}
\end{equation}
%

%%%
The ionization coefficients and photoionization heating rates are calculated by 
using the photon-conserving method \citep{Whalen_2006}.
The primary ionization rates $k_{{\rm ph},i}^{\rm p}$ ($i={\rm H, He}$, and ${\rm He^+}$) are given by
\begin{equation}
k_{{\rm ph},i}^{\rm p} = \int_{\nu_{\rm min}}^{\nu_{\rm max}} {\rm d}\nu~\dfrac{F_{\nu}}{h\nu}~ \sigma_{{\rm bf},i},
\end{equation}
where $\sigma_{{\rm bf},i}$ is the bound-free cross section.
Energetic electrons produced by primary ionization can further contribute to ionization of nearby atoms 
\citep[secondary ionization, e.g., ][]{shull_van_1985}.
The secondary ionization rates for species $j={\rm H}$, and ${\rm He}$ are
\begin{equation}
k_{{\rm ph},j}^{\rm s} = \sum_{i={\rm H, He}} \int_{\nu_{\rm min}}^{\nu_{\rm max}} {\rm d}\nu~\dfrac{F_{\nu}}{h\nu}
~\sigma_{{\rm bf}, j}  ~\Phi^{j}(E_{i},x_{\rm H^+}) ~\dfrac{x_{i}}{x_{j}},
\end{equation}
where $x_j$ is the abundance of species $j$, $\Phi^{j}(E_{i},x_{\rm H^+})$ is the fraction of secondary ionization of species $j$ 
per primary electron of energy $E_{i}\equiv h\nu - I_{i}$, 
and $I_i$ is the ground state ionization potential of the species $i$.
Thus, the photoionization rate is given by the sum of primary and secondary ionization rate.

The photoionization heating rate ($i={\rm H},~{\rm He}$, and ${\rm He^+}$) is calculated as
\begin{equation}
\Gamma_{i} = \int_{\nu_{\rm min}}^{\nu_{\rm max}} {\rm d}\nu~\dfrac{F_{\nu}}{h\nu} ~\sigma_{{\rm bf},i} ~E_h(E_{i},x_{\rm H^+}), 
\end{equation}
$E_h$ is the energy of a primary electron deposited as heat.
We adopt the functional forms of $\Phi^{\rm H}$, $\Phi^{\rm He}$, and $E_h$ \citep{ricotti_2002}.
Note that secondary ionization of ${\rm He}^{+}$ is negligible \citep{shull_van_1985}.

The radiation force caused by electron scattering and bound-free transition is calculated by
\begin{equation}
  f_{\rm rad} = \frac{nx_{\rm e}}{c}\int_{\nu_{\rm min}}^{\nu_{\rm max}}\sigma_{\rm es}
  F_{\nu}{\rm d}\nu + \frac{\Gamma^{\rm p}}{c},
  \label{f_rad}
\end{equation}
where $\Gamma^{\rm p}$ is the sum of the heating rates owing to primary ionization 
of ${\rm H,~He}$, and ${\rm He^+}$.

%%%
\subsection{Radiation spectra and anisotropy}
\label{sec:model}

To quantify the effect of radiative feedback (photoionization and heating), we take into account 
radiation spectra that depend on BH mass and accretion rate at each time step.
Following \cite{takeo_2019}, we adopt a multicolour blackbody spectrum \citep[e.g.,][]{kato+2008},
since the accretion rate onto the BH is always high enough ($\dot{M}_{\rm BH}/\dot{M}_{\rm Edd}\ga 10^{-2}$)
for the nuclear disc to be optically thick.
The specific radiation luminosity is give by 
\begin{equation}
L_{\nu}=2\int_{R_{\rm in}}^{R_{\rm out}}~{\rm d}R~2\pi R ~B_{\nu}[T_{\rm eff}(R)],
\end{equation} 
where $B_{\nu}(T_{\rm eff})$ is the blackbody intensity with an effective temperature of $T_{\rm eff}$, 
$R_{\rm in}(\simeq 1.1-1.7~R_{\rm Sch})$ and $R_{\rm out}(=10^4~R_{\rm Sch})$ 
are the locations of the disc inner and outer edge, respectively
(see more details in \citealt{takeo_2019}).
For both sub-Eddington and super-Eddington regimes, the radial profile of the effective temperature 
is given by 
\begin{equation}
T_{\rm eff}(R) = 2.5 \times 10^6~\K~ 
\left(\dfrac{M_{\rm BH}}{10^5~\msun}\right)^{-1/4} \left(\frac{R}{R_{\rm Sch}}\right)^{-1/2}  f^{1/8} \mathcal{F},
\label{disc_sol}
\end{equation}
where $f$ and $\mathcal{F}$ are functions that connects the standard and slim disc solution smoothly 
\citep[see the details in ][]{watarai_2006}.
Note that for a accretion rate of $\dot{m}_{\rm BH}(\equiv \dot{M}_{\rm BH}/\dot{M}_{\rm Edd})\gg 1$, 
$f\simeq 1$ and $\mathcal{F}\simeq 1$ inside the photon-trapping radius of $R_{\rm tr} \equiv \dot{m}_{\rm BH}R_{\rm Sch}$,
within which the advection timescale is shorter than the photon diffusion timescale.
Using this equation, the bolometric luminosity of radiation from a slim disc is expressed as
\begin{equation}
\frac{L}{L_{\rm Edd}} = 2\left[1+\ln \left(\frac{\dot{m}_{\rm BH}}{2}\right)\right],
\label{eq:Lmdot}
\end{equation}
for $\dot{m}_{\rm BH}>2$, otherwise $L/L_{\rm Edd}=\dot{m}_{\rm BH}$ \citep{watarai+00}.

We assume radiation field to be isotropic, i.e., $F_{\nu}(r=r_{\rm min}) = L_{\nu}/ (4\pi r_{\rm min}^2 )$
when the BH accretion rate is as low as $\dot{m}_{\rm BH} \leq 4$ (although the radiative flux from 
a geometrically thin disc has an angular dependence of $F_{\nu} \propto |\cos \theta|$).
This treatment gives us conservative conditions required for the transition to rapid accretion phases 
because highly anisotropic radiation does not prevents gas inflows from the equatorial region
which cover a large solid angle.
After the transition of gas inflow into hyper-Eddington accretion, the nuclear disc is likely to be 
geometrically thick with $H/R\simeq 0.3-0.5$ \citep{abramowicz_1988}, and thus radiation is emitted
preferentially towards the poles \citep{ohsuga+05,Jiang_2014,Sadowski_2015}.
Therefore, we assume anisotropic radiation with an opening angle of $\theta_{\rm rad}=45^{\circ}$
as a fiducial case after the transition (same as outflows).
In addition, we study a case with isotropic radiation even after the transition in \S\ref{sec:iso_rad} 
to conservatively discuss the stability of a hyper-Eddington phase because isotropic radiation can 
prevent the gas inflow from the equatorial region \citep[see also][]{takeo_2018}.

%%%
\subsection{initial and boundary conditions}
\label{sec:IBC}

To solve the basic equations, we employ spherical coordinates with a logarithmically spaced 
grid in the radial direction ($r_{\rm min}\leq r\leq r_{\rm max}$) and a uniformly-spaced grid 
in the polar direction ($0\leq \theta \leq \pi$; $\theta=\pi/2$ corresponds to the equator).
Since we study the gas dynamics over a wide range of spatial scales 
that cover the Bondi radius and includes its interior,
we adopt $(r_{\rm min},~r_{\rm max}) = (0.007~R_{\rm B}, ~6~R_{\rm B})$. 
The numbers of the grid cells are set to $(N_r,~N_{\theta}) = (60,120)$.
We note that the number of the radial grids in this study is lower than that in our previous work, 
where $N_r=100$ is set \citep[][]{takeo_2018}, in order to reduce the computational time but 
stably calculate mass flow with a high Mach number $\gg 1$ near the polar regions.

As our initial conditions, we set a neutral uniform and static (${\boldsymbol v}= 0$) gas cloud 
with a density $n_{\infty}$ and temperature $T_{\infty}=10^4~\K$ surrounding a BH with a mass of $M_{\rm BH}$.
The BH mass is assumed to be constant throughout the simulations.
We impose the absorption inner boundary conditions for the gas density, gas pressure, and 
velocity to be damped smoothly \citep[e.g.][]{kato+04}, and the free outer-boundary conditions for three 
components of the velocity.
The reflection symmetry with respect to the polar axis is imposed for 
non-radial components of the velocity.

To study the impact of mechanical feedback on BH accretion, 
we explore a wide range of the parameters to characterize outflow power: 
the outflow opening angle $45^{\circ} \leq \theta_{\rm out} \leq 80^{\circ}$, 
the mass loading factor $0.01 \leq \beta_{\rm out} \leq 10$, and
the outflow velocity $100~{\rm km~s^{-1}} \leq v_{\rm out} \leq 1,000~{\rm km~s^{-1}}$.
In Table \ref{models}, we summarize the models we study in this work.

%% Fig. 1 %%
\begin{figure}
\begin{center}
\includegraphics[width=8.3cm]{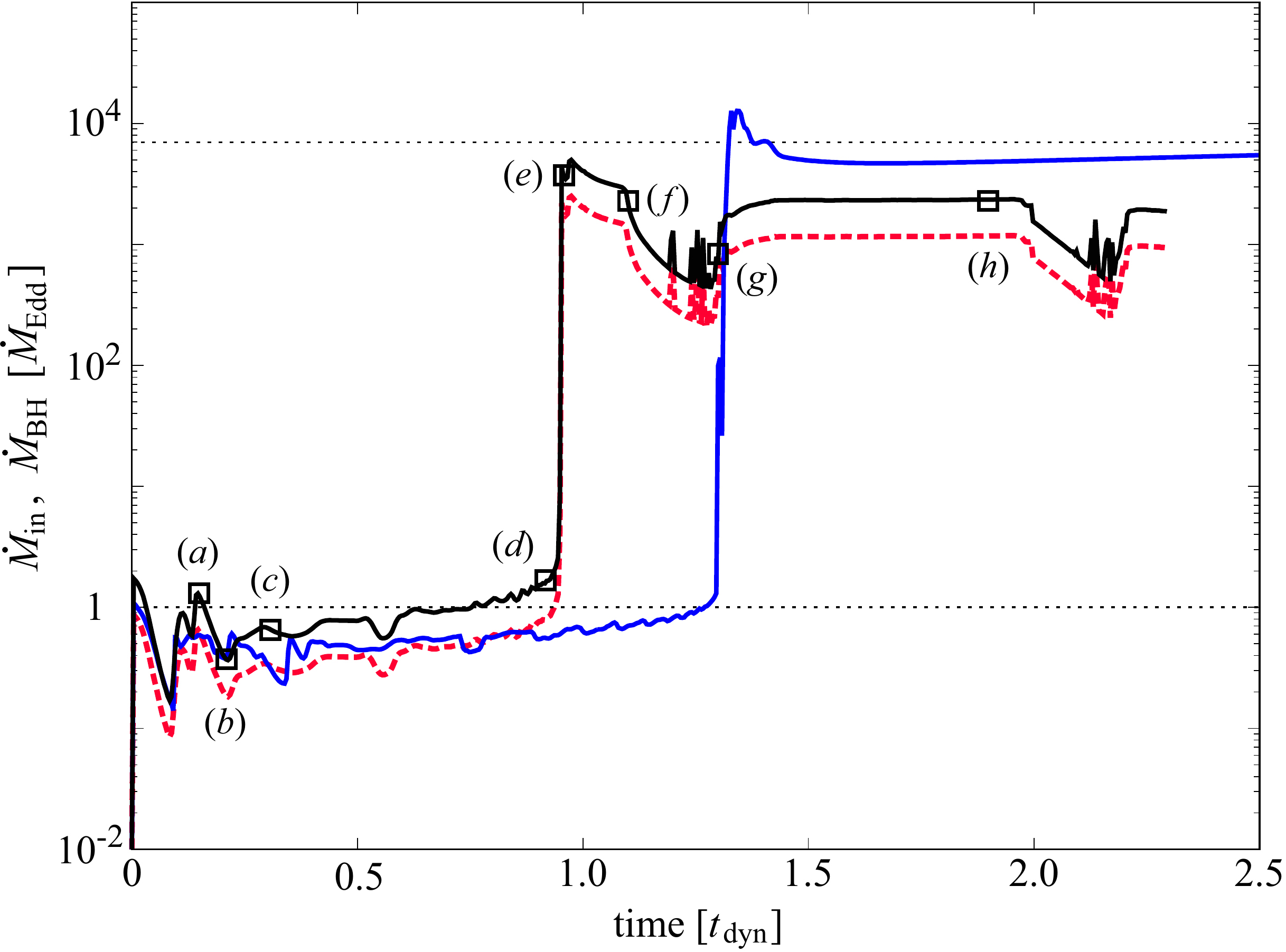}
\end{center}
\vspace{-2mm}
\caption{
Time evolution of mass inflow rates $\dot{M}_{\rm in}$ (in units of $\dot{M}_{\rm Edd}$) through $r=r_{\rm min}$
for a BH mass of $M_{\rm BH} = 10^5~\msun$ ($n_\infty = 10^5~\cc$ and $T_\infty=10^4~\K$ are
assumed as initial conditions) with both radiative and mechanical feedback (black, Model A45M5) and 
with radiative feedback alone (blue; Model RAD).
With mechanical feedback, the model parameters of outflow are set to $\theta_{\rm out}=45^{\circ}$, 
$\beta_{\rm out}=1$ and $v_{\rm out}=1,000~{\rm km~s^{-1}}$, and the BH accretion rate is shown by red dashed curve.
For both the cases, the BH accretion rate is limited below the Eddington value (lower dotted line) and
has a big jump at $t\simeq t_{\rm dyn}$ to a rate as high as the Bondi accretion rate (upper dotted line).
Open squares mark the eight epochs (phases {\it a-h}) at which the 1D radial profiles and 2D contours are shown in
Figs. \ref{cont_45d}, \ref{prof_45d}, \ref{cont_tran45d} and \ref{prof_tran45d}.
While mechanical feedback makes the transition to rapid accretion occur earlier, 
the mass inflow and BH accretion rate after the transition are reduced from the case with radiation alone 
by a factor of $\simeq 3$ and $6$, respectively.}
\label{t_md_angle}
\end{figure}

%%%%%%%%%
%   3. Results   %
%%%%%%%%%
\section{Results}
\label{sec:3}

\subsection{Super-Eddington accretion flows exposed to radiative and mechanical feedback}
\label{sec:angle}

%% Fig. 2 %%
\begin{figure}
\begin{center}
\includegraphics[width=8.3cm]{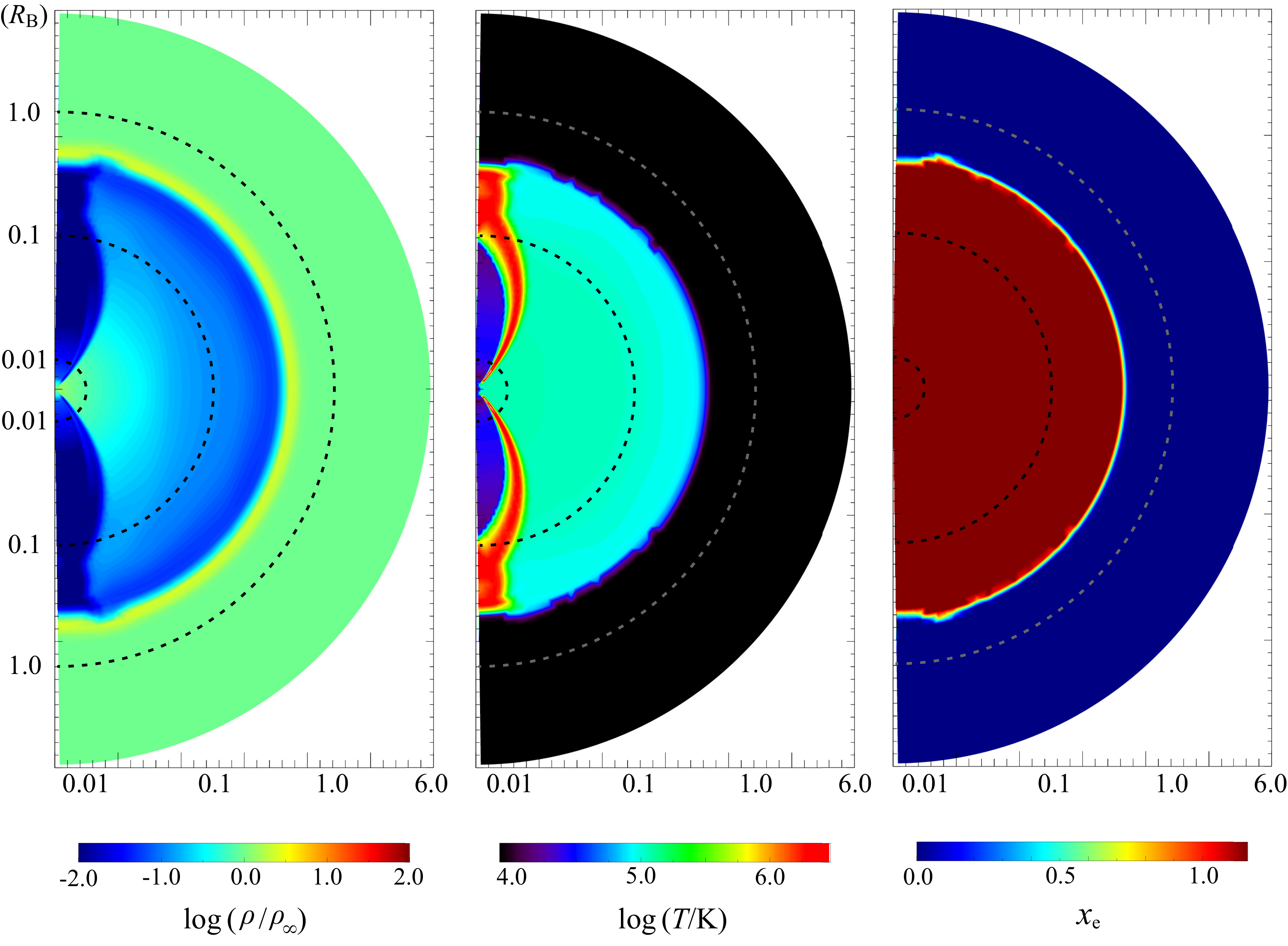}
\end{center}
\vspace{-2mm}
\caption{
Two-dimensional distribution of the gas density (left), temperature (middle), and ionization degree (right)
for Model A45M5 in the phase ($a$) shown in Fig. \ref{t_md_angle}.
The flow structure is divided into the bipolar outflowing polar region and the equatorial inflowing region,
both of which are embedded in an almost spherical ionized region.
}
\label{cont_45d}
\end{figure}

First, we discuss our fiducial case of a massive seed BH with $M_{\rm BH} =10^5~\msun$ surrounded by 
gas with density of $n_\infty = 10^5~\cc$.
The corresponding Bondi and Eddington accretion rates are $\dot{M}_{\rm B} = 15~\msunyr$ and 
$\dot{M}_{\rm Edd} =2.1 \times 10^{-3}~\msunyr$ (i.e., $\dot{M}_{\rm B}/\dot{M}_{\rm Edd}=7 \times 10^3$).
The model parameters for outflow injected at $r=r_{\rm min}$ are set to 
$\theta_{\rm out}=45^{\circ}$, $\beta_{\rm out}=1$ (i.e., $\dot{M}_{\rm BH}=\dot{M}_{\rm out}=0.5~\dot{M}_{\rm in}$),
and $v_{\rm out}=1,000~{\rm km~s^{-1}}$.
Radiation field is assumed to be isotropic when the accretion rate is a nearly or sub-Eddington value.
As seen below, the accretion flow turns into a hyper-Eddington phase, where radiation produced from the 
geometrically-thick accretion disc is supposed to be anisotropic.
In that case, therefore we consider anisotropic radiation field collimated with an opening angle of 
$\theta_{\rm rad}=45^{\circ}$ from the pole (see \S\ref{sec:supEddRM}).

Fig. \ref{t_md_angle} shows the time evolution of mass inflow rate $\dot{M}_{\rm in}$ through $r=r_{\rm min}$
and BH accretion rate $\dot{M}_{\rm BH}$ with both radiative and mechanical feedback (black and red curves, 
respectively; Model A45M5).
For reference, we also present the case with radiative feedback alone (blue curve; Model RAD) where isotropic radiation 
is assumed throughout the simulation.
With radiation feedback alone, mass accretion occurs episodically and the time-averaged rate is limited to 
$\lesssim \dot{M}_{\rm Edd}$.
The mass inflow rate becomes less time-dependent as time goes and abruptly rises to a rate as high as 
$\sim \dot{M}_{\rm B}$ at $t\simeq 1.3~t_{\rm dyn}$.
After the transition, the mass inflow rate becomes constant and thus the accretion flow settles down to 
a steady state, where ram pressure of the inflow overcomes momentum of incident radiation with luminosity 
of $L/L_{\rm Edd}\simeq 20$.
This transition behaviour has also been found in previous studies with different simulation codes
(see \citealt{inayoshi+16,takeo_2018,Toyouchi+2019} for more details).

%% Fig. 3 %%
\begin{figure}
\hspace{0mm}
\includegraphics[width=8.2cm]{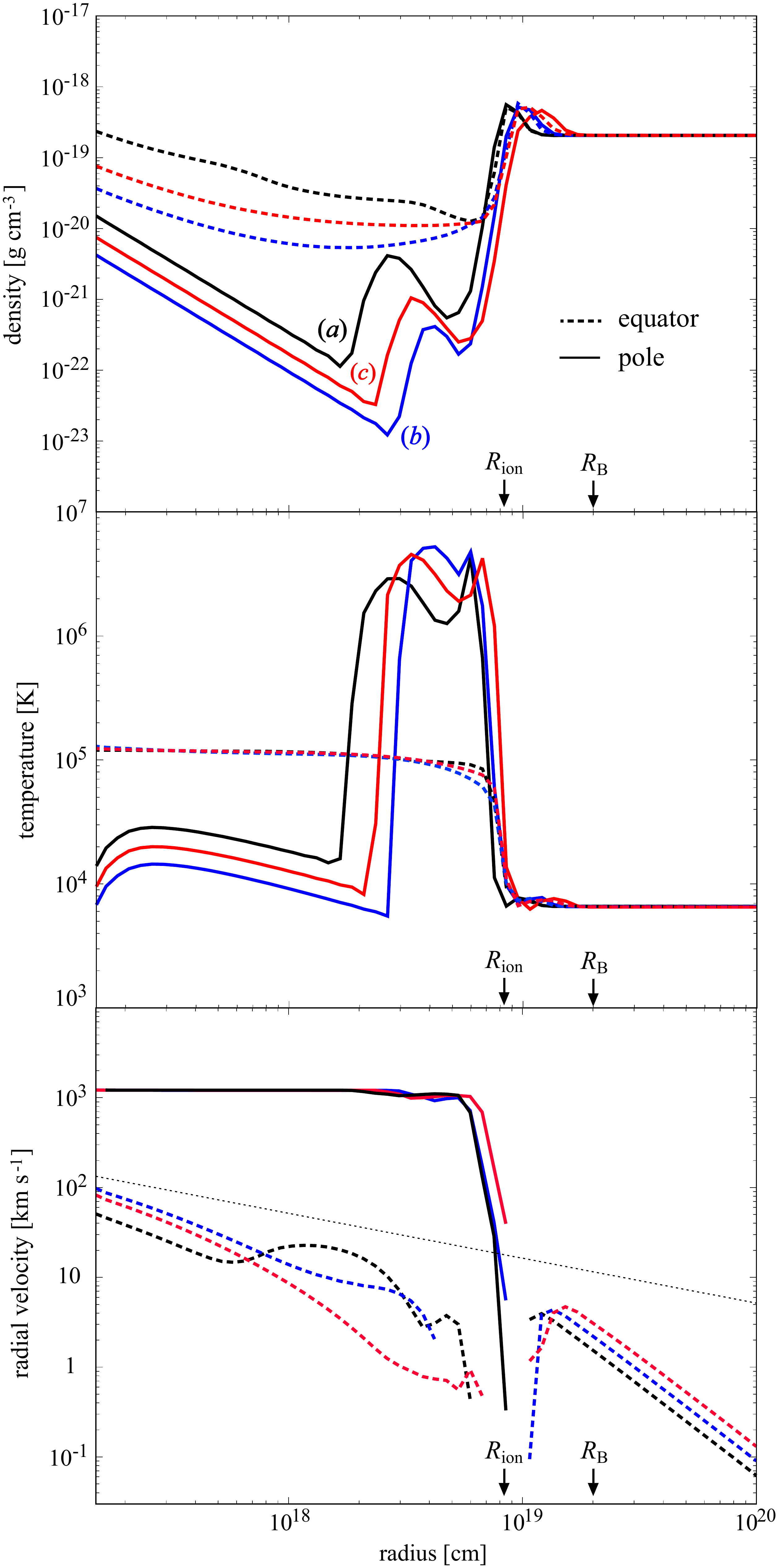}
\caption{
Radial structure of the gas density (top), temperature (middle), radial velocity (bottom) for Model A45M5
at three different epochs: $t/t_{\rm dyn} = 0.15$ (phase $a$; black), $0.21$ (phase $b$; blue), and $0.31$ (phase $c$; red).
The solid and dashed curves present the profiles along the pole ($\theta=0^\circ$) and equator ($\theta=90^\circ$).
In the bottom panels, only the outflow component along the pole and inflow component along the equator are shown.
The dotted line shows the free-fall velocity.
In the early stage, radiative and mechanical feedback heats and evacuates gas within the ionized region 
($R_{\rm ion}$), but do not affect the dynamics around the Bondi radius for neutral gas 
($R_{\rm B}$) substantially.
}
\label{prof_45d}
\end{figure}

%% Fig. 4 %%
\begin{figure*}
\begin{center}
\includegraphics[width=12.5cm]{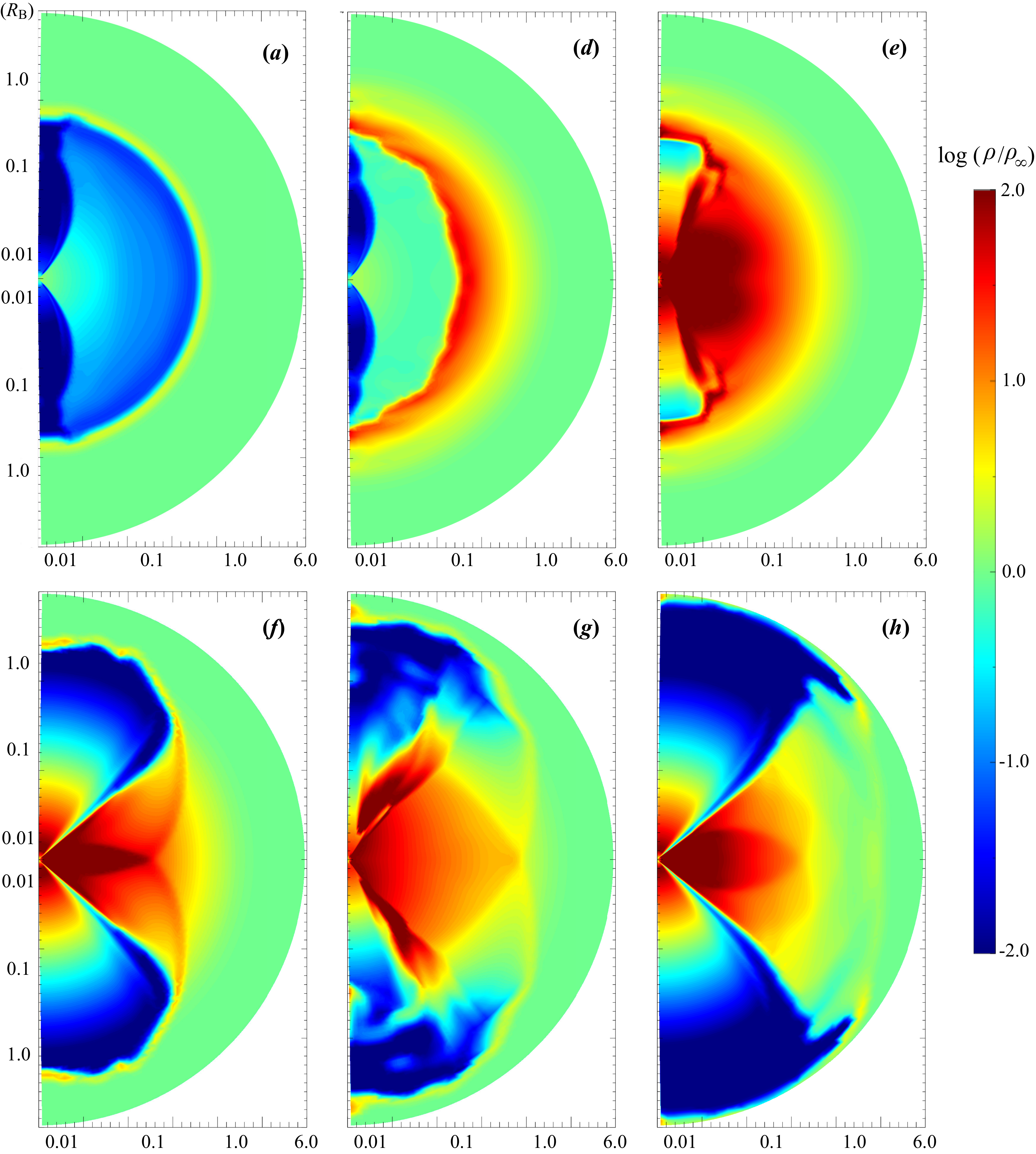}
\end{center}
\vspace{0mm}
\caption{
Two-dimensional distribution of the gas density for Model A45M5 at different elapsed times: 
({\it a}) $t/t_{\rm dyn} = 0.15$, ({\it d}) $0.92$, ({\it e}) $0.96$, ({\it f}) $1.1$, ({\it g}) $1.3$, and ({\it h}) $1.9$.
In the early stage (phases {\it a} and {\it d}), the ionized region is confined within the Bondi radius and 
both radiative and mechanical feedback do not affect the gas properties at larger radii.
In the transition epoch (phases {\it e} and {\it f}), the ionized region begins to collapse due to intense inflow of neutral gas 
through the equatorial region and then the mass inflow at rates of $\dot{m}_{\rm in} \gtrsim 2\times 10^3$ 
produces powerful bipolar outflows.
In the late stage (phases {\it g} and {\it h}), the bipolar outflows and emergent anisotropic radiation evacuate 
gas in the polar region completely, but a high inflow rate of $\langle \dot{m}_{\rm in}\rangle \sim 10^3$ is sustained.
}
\label{cont_tran45d}
\end{figure*}

In the case with both radiative and mechanical feedback, the overall behaviour of mass inflow rate is qualitatively similar, 
but there are remarkable differences from the case without outflows.
In fact, the mass inflow rate before the transition is twice higher (though the BH accretion rate is comparable) than those 
in the case with radiation alone, and the transition of accretion flows to a hyper-Eddington phase occurs earlier. 
This suggests that contrary to naive expectation, mechanical feedback {\it does} promote the transition 
(see discussion in \S\ref{sec:tran}).
After the transition when powerful outflows inject mechanical momentum of $\gtrsim 30~L_{\rm Edd}/c$ into the polar region, 
the mass inflow and BH accretion rate are reduced by a factor of $\simeq 3$ and $6$, respectively.
However, the reduced accretion rate is still as high as $\sim 10^3~\dot{M}_{\rm Edd}$.

In what follows, we show the detailed structure of accretion flows in our fiducial case (Model A45M5) 
and describe how mechanical feedback affects the accretion flow from the Bondi scale in more details, 
dividing the entire episode into two epochs before and after the transition.

%% Fig. 5 %%
\begin{figure*}
\includegraphics[width=17.3cm]{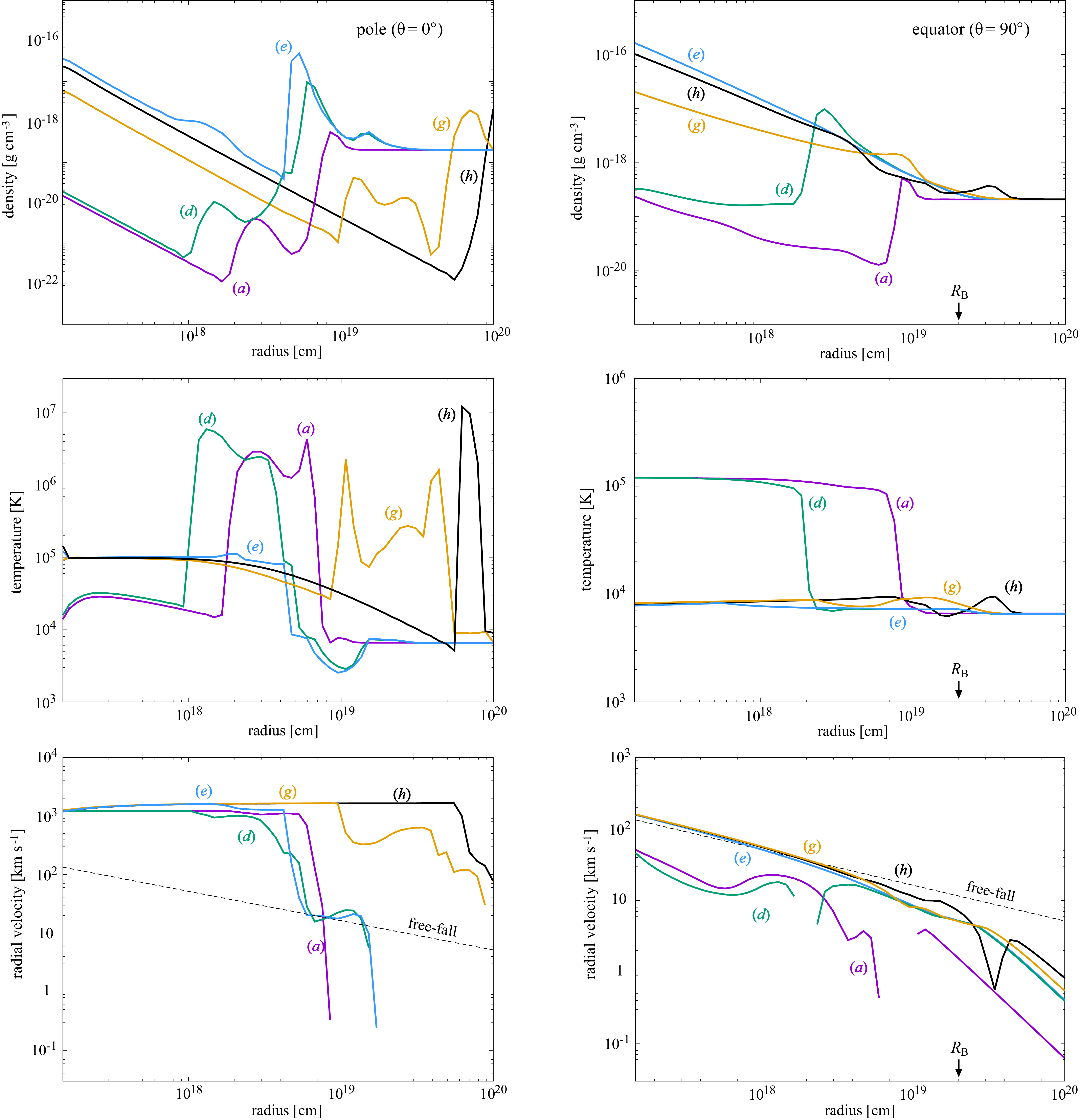}
\vspace{2mm}
\caption{
Radial structure of the gas density (top), temperature (middle), radial velocity (bottom) for Model A45M5
along the pole ($\theta=0^\circ$; left panels) and equator ($\theta=90^\circ$; right panels).
In each panel, we show the profiles at different epochs during and after the transition:
$t/t_{\rm dyn} = 0.15$ (phase {\it a}; purple), 
$0.92$ (phase {\it d}; green), 
$0.96$ (phase {\it e}; blue),
$1.3$ (phase {\it g}; yellow),
and $1.9$ (phase {\it h}; black).
In the bottom panels, only the outflow component along the pole and inflow component along the equator are shown.
After the transition, the equatorial accretion flow settles to the isothermal Bondi solution with $T\simeq 8000~\K$, 
whereas the outflow head continues to move outward, leading to the break out of the ambient gas cloud.
}
\label{prof_tran45d}
\end{figure*}

%%%
\subsubsection{Episodic accretion owing to radiative feedback}

In Fig. \ref{cont_45d}, we show the two-dimensional distribution of gas density (left), temperature (middle) 
and ionization degree (right) at $t/t_{\rm dyn} = 0.15$ (phase {\it a} in Fig. \ref{t_md_angle}).
Fig. \ref{prof_45d} presents the radial profiles of gas density (top), temperature (middle), radial velocity (bottom)
along the pole (solid) and equator (dashed) at three different epochs in the early stage; 
$t/t_{\rm dyn} = 0.15$ (black; phase $a$), $0.21$ (blue; phase $b$), and $0.31$ (red; phase $c$).
In the bottom panel, we show only the outflow component along the pole and 
inflow component along the equator.

In the early stage, ionizing radiation emitted from the central accreting BH heats the surrounding gas
and forms a nearly spherical low-density cavity until the ionization front reaches $\simeq 0.5~R_{\rm B}$.
In the ionized region with $T\simeq 10^5~\K$, the outflow associated with mechanical feedback blows 
the gas outward and creates even lower-density cavities near the poles where gas cools owing to adiabatic expansion.
The outflow collides with the surrounding gas and forms shocked gas with temperature of $\sim 10^{6-7}~\K$,
which corresponds to the post-shock temperature of the pre-shock outflowing matter at a velocity of $1,000~{\rm km~s}^{-1}$.
Although the outflow velocity is ten times higher than the escape velocity from the inner-most grid, 
the mechanical power ($\dot{P}_{\rm out}\simeq 0.015~L_{\rm Edd}/c$) is not strong enough to 
break the surrounding medium.

In the equatorial region, where the mechanical momentum of outflows is not injected,
the gas begins to accrete from the sonic point for the ionized gas, 
$R_{\rm B,ion} \equiv GM_{\rm BH}/c_{\rm ion}^2 \sim 2\times 10^{18}~{\rm cm}$, 
where $c_{\rm ion}$ is the sound speed of the ionized gas.
On the contrary, gas outside $R_{\rm B,ion}$ is pushed outward by the negative pressure gradient force.
As the ionized gas is depleted in the cavity and the outward gas pressure force decreases, 
a density bump forms just inside the edge of the ionized region near the equator. 
Finally, the density bump exerted by the inward gas pressure (i.e. $dp/dr > 0$) falls into the central BH
(phase $c$).
However, radiative and mechanical feedback regulates further mass accretion, leading to episodic 
accretion several times before the transition occurs.
We note that this oscillatory behaviour is caused by the same mechanism as studied in previous work
\citep[e.g.][]{Ciotti_Ostriker_2001,Milosavljevic_2009a,Park_Ricotti_2011,Park_Ricotti_2012,inayoshi+16}.

%%%
\subsubsection{Transition to hyper-Eddington accretion with powerful bipolar outflows}
\label{sec:supEddRM}

In Figs. \ref{cont_tran45d} and \ref{prof_tran45d}, we present the two-dimensional density distribution 
and the radial profiles of physical quantities at the early stage (phase $a$) and several different epochs 
in the late stage (phases $d$-$h$) including the transition and rapid accretion phase.

As shown in Fig. \ref{t_md_angle}, the accretion behaviour becomes less episodic after several oscillations.
In the quasi-steady state ($0.3 \lesssim t/t_{\rm dyn}\lesssim 0.8$), the expansion of the ionized region halts and 
its size reaches a maximum value of $R_{\rm ion}\simeq 8\times 10^{18}~{\rm cm}$, which is smaller than the Bondi radius.
Since ionizing radiation does not heat and prevent gas inflows from $R_{\rm B}$, neutral gas is piled up at 
$R_{\rm ion} \lesssim r \lesssim R_{\rm B}$ (phase $d$ in Fig. \ref{cont_tran45d}).
As the dense shell increases its mass and pushes the hot gas inwards, the ionized region gradually 
shrinks from the equatorial region and finally collapses (phase $e$ in Fig. \ref{cont_tran45d}).
The density cavity near the equator is totally filled by strong inflows of neutral warm gas 
(see phase $e$ in Fig. \ref{cont_tran45d} and right panels of Fig. \ref{prof_tran45d}),
leading to a big jump of inflow rate to $\dot{m}_{\rm in} (\equiv \dot{M}_{\rm in}/\dot{M}_{\rm Edd})\approx 3.6 \times 10^3$.
In the equatorial region, the accretion flow settles to an isothermal Bondi solution with $T\approx 8000~\K$.
On the contrary, since the intense mass inflow produces powerful outflows towards the poles, 
collapsing dense shell from the polar regions is blown away before reaching $r=r_{\rm min}$
(see left panels of Fig. \ref{prof_tran45d}).

After this abrupt transition, the inflowing region gradually becomes larger and starts to overlap with 
the outflowing region, suppressing mass inflows (phase {\it f} in Fig. \ref{cont_tran45d}). 
As the feedback strength becomes relatively weaker, a dense clump forms near the interface between 
the inflow and outflow region and falls into the central BH.
However, since the inward ram pressure of the inflowing gas is not strong enough to overcome 
the momentum input by the outflow, the clump is blown away and thus the inflow rate sharply drops to 
$\dot{m}_{\rm in} \simeq 6\times 10^2$. 
The episodes of clump formation and ejection are repeated but the mass inflow rate settles to
a high value between $6\times 10^2 \lesssim \dot{m}_{\rm in} \lesssim 2.3\times 10^3$ (phase {\it g} in Fig. \ref{cont_tran45d}).
As a result, the polar region is clearly evacuated and thus the central radiating source is not wholly 
covered by neutral gas as found in the case without mechanical feedback \citep{takeo_2018}.
In the final stage of our simulation (phase $h$),
the bipolar outflowing region almost totally breaks out the ambient gas cloud in the polar region.
The gas temperature is shock heated up to $\sim 10^7~\K$ at the edge of the density cavity,
and the outflow velocity is $\approx 1,000~{\rm km~s^{-1}}$ at all radii (left panels of Fig. \ref{prof_tran45d}).

%%%
\subsection{Effects of outflow strength and radiation anisotropy after the transition}

\subsubsection{Higher mass loading factor: $\beta_{\rm out}=10$}
\label{sec:highbeta}

In the above section, we set the mass loading factor to $\beta_{\rm out}=1$ through the entire simulation.
However, theoretical and numerical studies suggest that the mass loading factor could be larger than 
unity after the transition to hyper-Eddington accretion (see \S\ref{sec:outpara}).
Thus, we investigate a model with $\beta_{\rm out}=10$ (Model A45M5-B10)
to see the impact on the flow structure and differences from the fiducial model (Model A45M5; $\beta_{\rm out}=1$).

%% Fig. 6 %%
\begin{figure}
\begin{center}
\includegraphics[width=8.3cm]{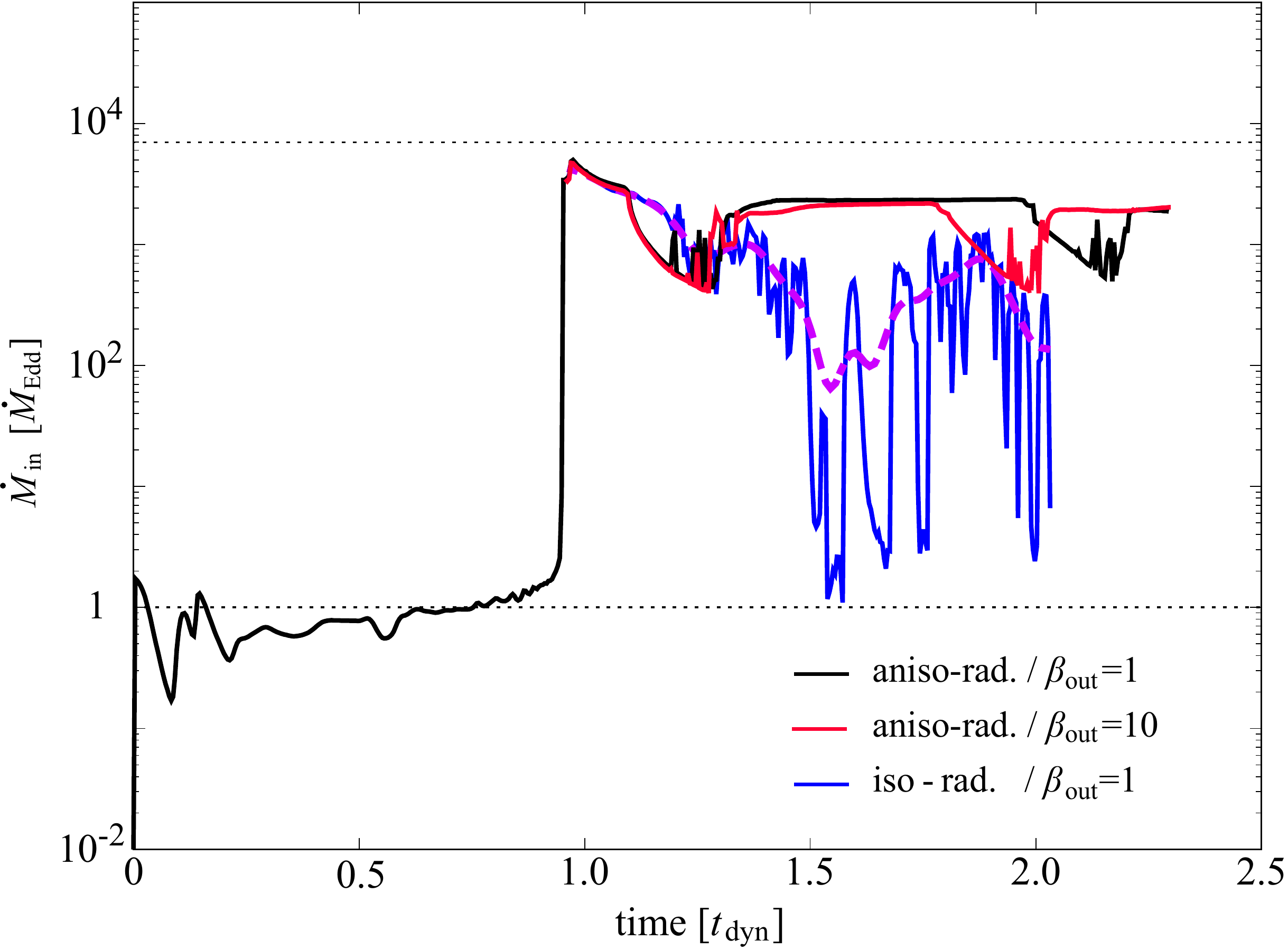}
\end{center}
\vspace{0mm}
\caption{
Time evolution of mass inflow rates $\dot{M}_{\rm in}$ for Model A45M5 (black; $\beta_{\rm out}=1$), 
A45M5-B10 (blue; $\beta_{\rm out}=10$ after the transition), and A45M5-B1-iso (blue; $\beta_{\rm out}=1$).
For Model A45M5-B1-iso, radiation field is assumed to be isotropic throughout the simulations, 
otherwise anisotropic radiation with an opening angle of $\theta_{\rm rad}=45^\circ$ is assumed after the transition.
In the cases anisotropic radiation, the mass inflow rate is as high as $\dot{m}_{\rm in}\simeq 2\times 10^3$,
nearly independent of the mass loading factor.
With isotropic radiation, the mass inflow rate is highly episodic, but the time-averaged value is 
$\langle \dot{m}_{\rm in}\rangle \sim 800~\dot{M}_{\rm Edd}$ (purple).
}
\label{t_md_acc_out}
\end{figure}

In Fig. \ref{t_md_acc_out}, we show the time evolution of mass inflow rates for $\beta_{\rm out}=1$ (black) 
and $\beta_{\rm out}=10$ (red).
For the two cases, the time-averaged mass inflow rate is almost equivalently as high as 
$\dot{m}_{\rm in} \simeq 2\times 10^3$, and their evolution is qualitatively similar.
This clearly shows that the choice of the mass loading factor does not affect the mass inflow rate 
as long as outflowing matter reverses more than 50\% of the inflowing matter, i.e., $\beta_{\rm out}\gtrsim 1$.
One might make a misinterpretation that a higher mass loading factor leads to suppression 
of BH accretion because the BH accretion seems lower.
However, the reduction of BH accretion rate is not caused by mechanical feedback, but is by definition of 
$\dot{M}_{\rm BH}=\dot{M}_{\rm in}/(1+\beta_{\rm out})$.

%% Fig. 7 %%
\begin{figure}
\begin{center}
\includegraphics[width=8.3cm]{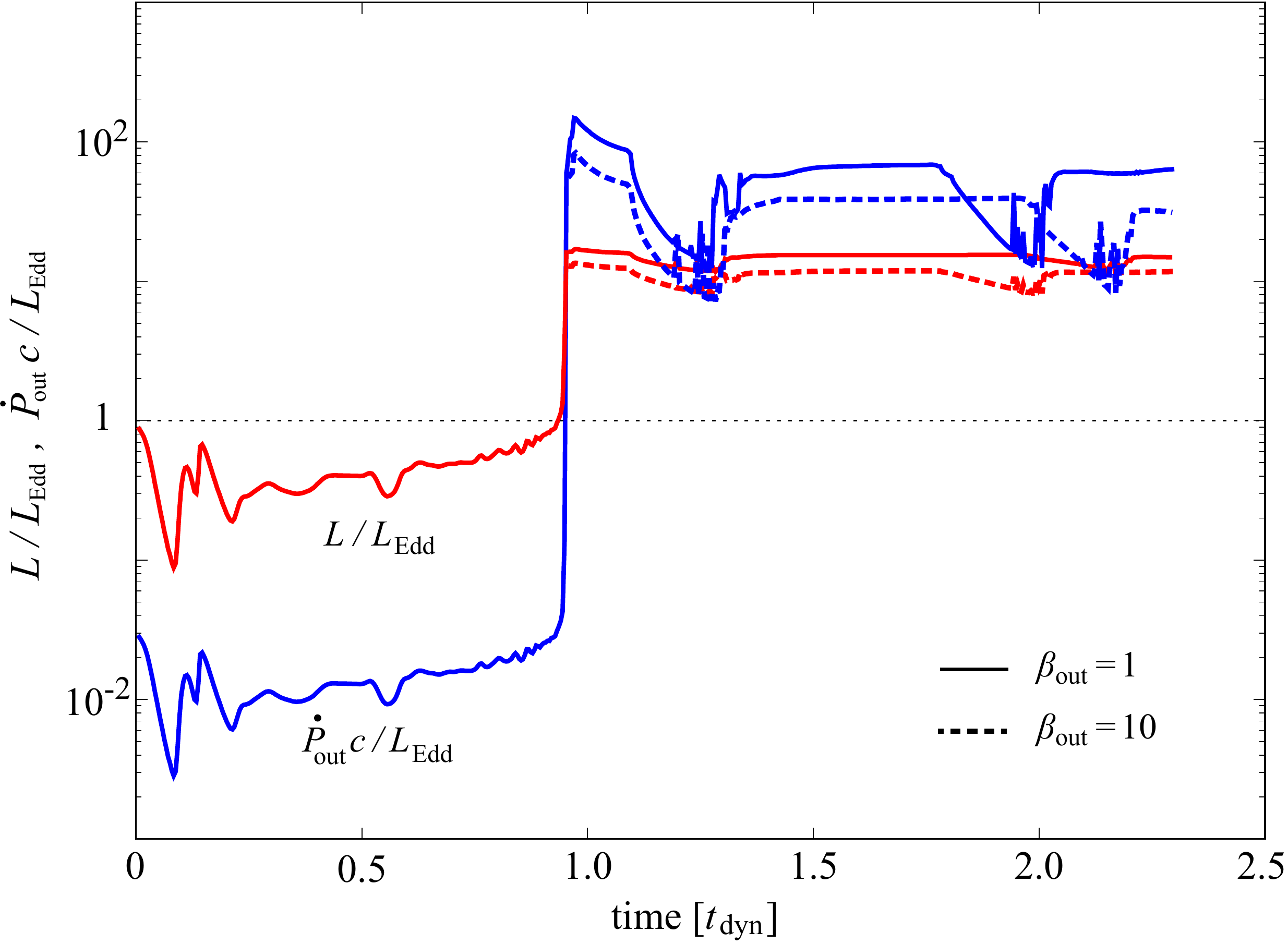}
\end{center}
\vspace{0mm}
\caption{
Time evolution of the radiative luminosity $L_{\rm rad}$ (red) and mechanical momentum input 
$\dot{P}_{\rm out}$ (dashed) normalized by the Eddington values for Model A45M5 (solid; $\beta_{\rm out}=1$), 
and A45M5-B10 (dashed; $\beta_{\rm out}=10$).
While radiative feedback dominates before the transition, the gas dynamics after the transition
is mainly determined by mechanical feedback associated with hyper-Eddington accretion.
}
\label{t_md_output}
\end{figure}

In Fig. \ref{t_md_output}, we present the radiative luminosity and mechanical momentum input 
(red and blue curves, respectively) for $\beta_{\rm out}=1$ (solid) and $\beta_{\rm out}=10$ (dashed), 
each of which is normalized by the Eddington value.
Before the transition, the strength of mechanical feedback is as low as $\sim 0.01~L_{\rm Edd}/c$,
whereas the radiation luminosity is comparable to $\simeq 0.3~L_{\rm Edd}$.
This shows that gas inflow is prevented mainly by radiative feedback in the early phase. 
On the contrary, the power of outflows dramatically rises up to $\sim (30-100)\times L_{\rm Edd}/c$ 
after the transition, although the radiation luminosity is saturated at $\sim 20~L_{\rm Edd}$ 
for the two cases because $L_{\rm rad}/L_{\rm Edd}\sim {\rm ln} \left(\dot{m}_{\rm BH} \right)$.
Thus, the flow structure is mainly determined by mechanical feedback in the later phase.

%%%
\subsubsection{Isotropic photospheric radiation field}
\label{sec:iso_rad}

%% Fig. 8 %%
\begin{figure*}
\begin{center}
\includegraphics[width=13.3cm]{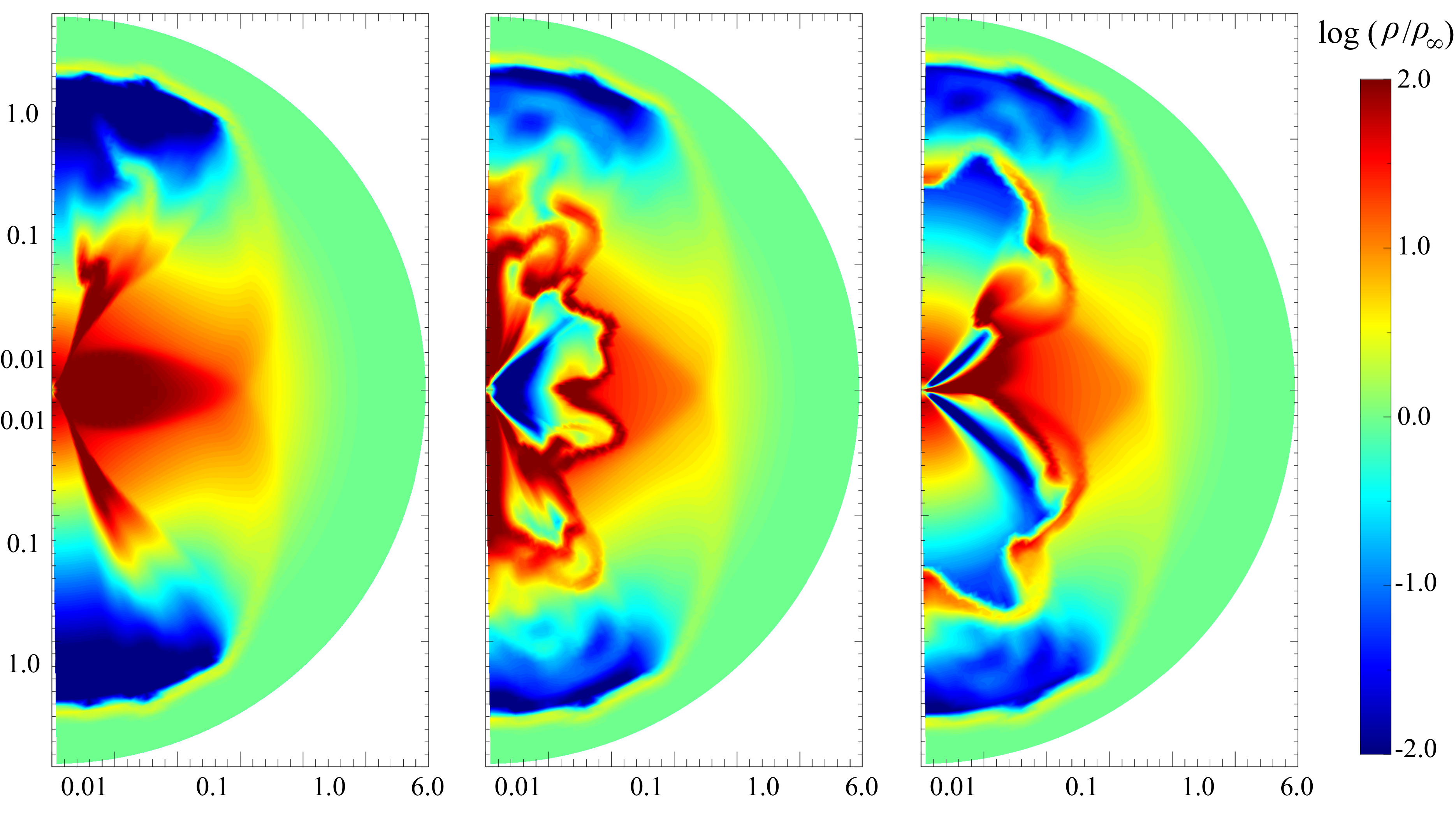}
\end{center}
\vspace{0mm}
\caption{
Two-dimensional distribution of the gas density for Model A45M5-B1-iso at different elapsed times of
$t/t_{\rm dyn} = 1.20$ (left), $1.28$ (middle), and $1.30$ (right).
Unlike the anisotropic radiation case, dense gas around the equatorial region is blown outward by radiative feedback 
associated with hyper-Eddington accretion at a time-averaged rate of $\dot{m}_{\rm in}\simeq 800$ (and $\dot{m}_{\rm BH}\simeq 400$).
}
\label{cont_BBdlt45d}
\end{figure*}

So far, we assume radiation flux emitted from a geometrically-thick disc after the transition
to be anisotropic collimated within an opening angle of $\theta_{\rm rad}=45^\circ$.
However, radiative flux to the equatorial region is not zero but would be a significant fraction of $L_{\rm Edd}$.
Presumably, the non-collimated radiation component would be emitted from the photosphere located at 
the outer-most disc radius or photon-trapping radius of $\sim O(10^4)~R_{\rm Sch}$.
Moreover, in such a high accretion rate, even the polar funnel regions become optically thick in the nuclear region
and radiation is effectively trapped and advected towards the BH together with inflowing matter.
As a result, most of the radiation energy is transported outward in the diffusion process \citep{Jiang_2019}.
For these reasons, we here simply assume that even after the transition, the radiation field is still isotropic
with a diluted blackbody spectrum of an effective temperature of 
$T_{\rm eff} = (L/4\pi R_{\rm tr}^2 \sigma_{\rm SB})^{1/4}$ (Model A45M5-B1-iso),
where $\sigma_{\rm SB}$ is the Stefan-Boltzmann constant and the photospheric size is approximated to be $\simeq R_{\rm tr}$,
although the realistic radiation anisotropy would be somewhere between this case and previous case 
discussed in Section \ref{sec:supEddRM}.

We show the time evolution of mass inflow rate (blue) and the rate averaged over a longer timescale (purple)
for the isotropic radiation case in Fig. \ref{t_md_acc_out}, and present two-dimensional distribution of the gas density 
after the transition in Fig. \ref{cont_BBdlt45d}.
Similarly to the anisotropic radiation case, high density clumps form and accrete onto the centre 
through the polar regions, leading to a high inflow rate of $\dot{m}_{\rm in} \sim O(10^3)$.
Since radiation isotropically ionizes and heats the accreting matter,
the flux preferentially propagates towards the equatorial region, where the density is relatively lower 
than that in the polar regions.
As a result, dense gas near the equatorial region is pushed outward and thus the total mass inflow 
is reduced by 1-2 orders of magnitude (middle panel).
Although powerful outflows evacuate the equatorial region and reduces the mass inflow rate, 
inflowing gas accretes again through the equatorial region with the aid of inward force exerted 
by positive gas pressure-gradient in the surrounding gas.
Because of the symmetry breaking, the mass inflow occurs episodically but the time-averaged rate 
over $1.2 \leq t/t_{\rm dyn} \leq 2.0$ is $\dot{m}_{\rm in}\simeq 8.1 \times 10^2$ (see purple curve in Fig. \ref{t_md_acc_out}).

%%%
\subsection{The effects of outflow parameters on the transition}
\label{sec:out_para}

%% Fig. 9 %%
\begin{figure}
\begin{center}
\includegraphics[width=8.0cm]{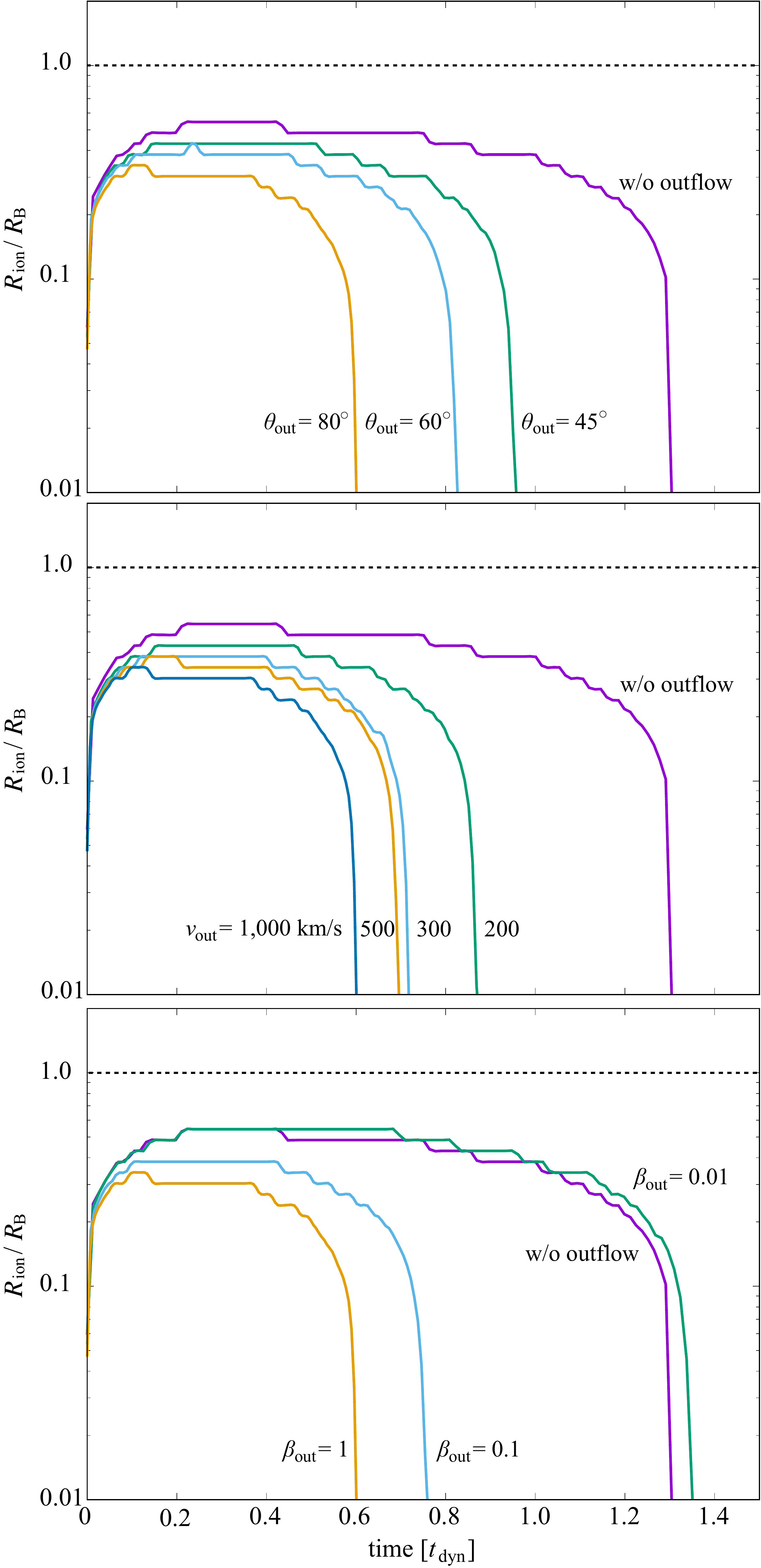}
\end{center}
\vspace{-2mm}
\caption{
{\it Top panel}: Time evolution of the ionization radii 
for the cases with radiative 
feedback alone and with mechanical feedback injecting outflows having an opening angle of 
$\theta_{\rm out}=45^\circ$, $60^\circ$ and $80^\circ$. 
{\it Middle panel}: Same as the model with $\theta_{\rm out}=80^\circ$ but for different outflow velocities;
$v_{\rm out}=200$, $300$, $500$ and $1,000~{\rm km~s}^{-1}$.
{\it Bottom panel}: Same as the model with $\theta_{\rm out}=80^\circ$ but for different mass loading factors;
$\beta_{\rm out}=0.01$, $0.1$ and $1.0$.
As the outflow is stronger (i.e., a wider opening angle, higher velocity and higher mass loading factor), 
the transition to rapid accretion tends to occur in a shorter dynamical timescale.
}
\label{t_rhii}
\end{figure}

Since the outflow launching mechanisms and outflow properties are still poorly understood, 
we explore the dependence of the transition behaviour on the choice 
of model parameters; the outflow opening angle $\theta_{\rm out}$, the mass loading factor 
$\beta_{\rm out}$ and the outflow velocity $v_{\rm out}$ (Model A60M5 $-$ B001M5; see Table \ref{models}).
In Fig. \ref{t_rhii}, we track the time evolution of the size of the ionized region $R_{\rm ion}$
and find that the transition to rapid accretion tends to occur earlier as the outflow is stronger,
i.e., larger values of $\theta_{\rm out}$, $\beta_{\rm out}$ and $v_{\rm out}$.

The reason why mechanical feedback promotes the hyper-Eddington transition is because 
powerful outflows reduce the BH accretion and mass inflow rate in the following ways.
When the outflow is launched into a large opening angle with a higher velocity, mass inflow is allowed only through 
the equatorial region with a solid angle of $\Delta \Omega_{\rm in} \propto \cos \theta_{\rm out}$,
otherwise strong outflows injected at the bottom reverse inflowing gas.
Furthermore, a fraction $\beta_{\rm out}/(1+\beta_{\rm out})$ of the mass inflow rate is loaded into outflows, 
which results in a lower BH accretion rate; $\dot{M}_{\rm BH} \propto \Delta \Omega_{\rm in}/(1+\beta_{\rm out})$.
Therefore, ionizing radiation attenuated by inefficient BH feeding creates a smaller ionized region 
that collapses within a shorter dynamical timescale as seen in Fig. \ref{t_rhii}.

Recently, \cite{Regan+2019} studied the impact of bipolar jets launched from an accreting seed BH on the gas dynamics
using cosmological simulations, which do not consider radiative feedback due to ionizing photons.
They found that the powerful jets evacuate the surrounding gas in the central $\sim 0.1$ pc ($\lesssim R_{\rm B}$)
and lead to highly episodic mass accretion at rates ranging $10^{-3}\lesssim \dot{m}_{\rm BH}\lesssim 1$.
However, their simulations do not see a transition into rapid accretion, even though neither the jets nor radiation prevents 
gas supply from the halo scales down to the Bondi scale.
This discrepancy from our result may be because the injected feedback strength is different by orders of magnitude.
In fact, they consider a constant value for the jet efficiency $\eta_{\rm jet}\sim 0.64 ~[\equiv L_{\rm out}/(\dot{M}_{\rm BH}c^2)]$, 
assuming that the innermost disc can be a magnetically arrested disc \citep{Tchekhovskoy+2011}, 
where the jet efficiency depends on the BH spin $a_{\rm BH}$ and can be described as $\eta_{\rm jet}\simeq 1.3~a_{\rm BH}^2$ 
\citep{Tchekhovskoy+2015}.
Since the jet efficiency is rewritten as $\eta_{\rm jet}=0.5\beta_{\rm out}(v_{\rm out}/c)^2$, 
the mass-loading factor adopted in \cite{Regan+2019} is as high as $\beta_{\rm out}\simeq 100~(10^4)$ 
for $v_{\rm out}=0.1c~(0.01c)$.
Such a high mass loading factor leads to a significant reduction of BH accretion rate, 
i.e., $\dot{M}_{\rm BH}\sim \dot{M}_{\rm in}/(1+\beta_{\rm out})$, and limit the rate below the Eddington value
even if the mass inflow rate is a super/hyper-Eddington value independent of $\beta_{\rm out}$ 
(see also discussion in \ref{sec:highbeta} and Fig. \ref{t_md_acc_out}).

%%%%%%%%%%%
%   4. Discussion    %
%%%%%%%%%%%
\section{Discussion}
\label{sec:4}

%%% 
\subsection{Revised transition criterion}
\label{sec:tran}

\begin{figure}
\begin{center}
\includegraphics[width=8.0cm]{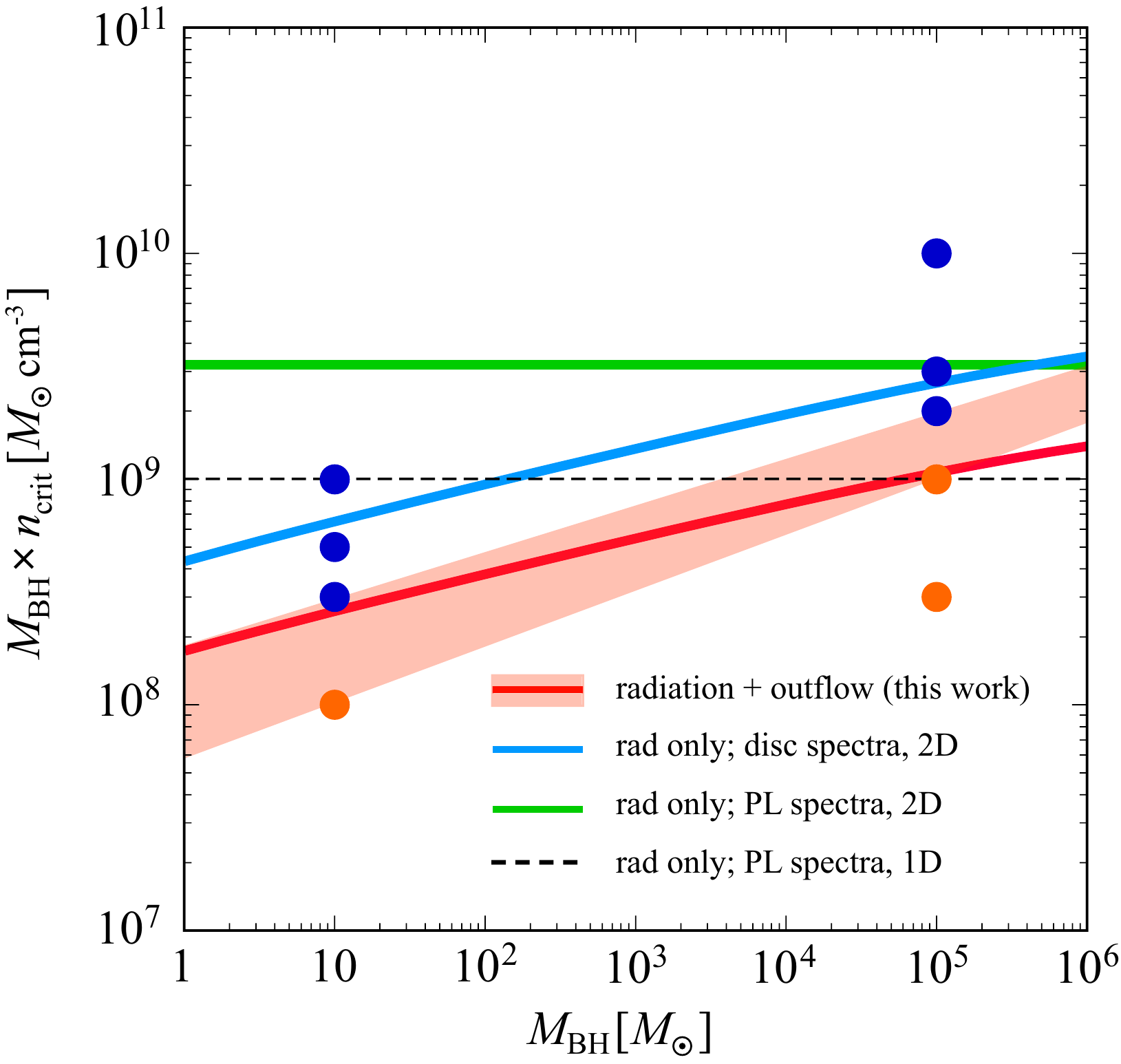}
\end{center}
\vspace{-2mm}
\caption{
Summary of our simulation results for $M_{\rm BH} =10$ and $10^5~\msun$. 
The outflow opening angle is set to $\theta_{\rm out}=80^{\circ}$, 
the mass loading factor is $\beta_{\rm out}=1$, 
and the outflow velocity is $v_{\rm out}=1,000~{\rm km~s^{-1}}$.
Each symbol indicates whether the final result is hyper-Eddington accretion (blue circle), 
or episodic accretion at the rate of $\lesssim \dot{M}_{\rm Edd}$ (orange circle).
The red shaded region marks the boundary between the two accretion modes obtained from the simulation data.
Each solid curve shows the transition criterion derived by the analytical argument in \S\ref{sec:tran} that takes into account
both radiative and mechanical feedback (red; see Eq. \ref{n_crit}), radiative feedback alone assuming disc spectra
(blue) and power-law spectra (green), respectively.
The critical density required for hyper-Eddington accretion with outflows is reduced by a factor of $\sim 2-4$ 
from that with radiative feedback alone nearly independent of BH mass, and is also lower than 1D simulations estimated 
\citep[dashed line][]{inayoshi+16}.
}
\label{survey}
\end{figure}

In our previous studies, we quantify the conditions required for the transition to rapid accretion
only when the accreting gas is exposed to ionizing radiation associated with BH feeding 
\citep{inayoshi+16,takeo_2018,takeo_2019}, where the conditions are characterized by
the dimensionless Bondi accretion rate\footnote{
The exact value of critical rate depends on the radiation spectral model and has a scaling relation of 
$\propto \langle h\nu \rangle ^{-5/9}$, where $\langle h\nu \rangle$ is the mean energy of ionizing 
radiation \citep{takeo_2019}.}; $\dot{M}_{\rm B}/\dot{M}_{\rm Edd}\gtrsim 500$.
We here extend this criterion for the transition to the case with mechanical feedback.
In Fig. \ref{survey}, we summarize the results for two different BH masses of $M_{\rm BH}= 10$ and $10^5~\msun$
with several different values of $n_{\infty}$,
where the outflow model parameters are set to $\theta_{\rm out} = 80^{\circ}$, $\beta_{\rm out}=1$, and 
$v_{\rm out}=1,000~{\rm km~s^{-1}}$ (Model T34M5 $-$ T17M1; see Table \ref{models}).
In order to save the computation time, we adopt $\theta_{\rm out} = 80^{\circ}$ because the transition, if any,  
occurs in a shorter timescale (see discussion in \S\ref{sec:out_para}).
For each BH mass, the transition to hyper-Eddington accretion is found in cases with higher ambient gas density 
(blue circles), while the transition is not seen in cases with lower density within $\sim 5~t_{\rm dyn}$ (orange circles).
The red shaded region marks the boundary between the two accretion modes under mechanical feedback.
We also overlay the critical conditions obtained by 1D RHD simulations \citep[][black line]{inayoshi+16}
and 2D RHD simulations with power-law spectra and disc multi-colour blackbody spectra 
(green and blue region, respectively; \citealt{takeo_2019}).
The results show that mechanical feedback reduces the critical density by a factor of 
$\simeq 2-4$ nearly independently of BH mass.

%% Fig. 10 %%

The critical conditions for the hyper-Eddington transition is evaluated in the following analytical argument,
comparing the Bondi radius and the ionization radius \citep{inayoshi+16}.
We approximate the size of ionized regions by the Str\"{o}mgren radius, where 
the number flux of ionizing photons for disc radiation spectra is estimated as
\begin{equation}
\dot{N}_{\rm ion} = \int_{\nu_{\rm min}}^{\nu_{\rm cut}}~{\rm d}\nu~ \dfrac{L_{\nu}}{h\nu}
\approx  2 \times 10^{46}~{\rm s^{-1}}~ M_{\rm BH}^{1.3}~\dot{m}_{\rm BH}^{0.75},
\label{ndot_disc}
\end{equation}
where the above expression is valid at $1< M_{\rm BH}/\msun <10^6$ and $10^{-2}< \dot{m}_{\rm BH} < 10$.
We note that the cutoff frequency for the integration is set to $h\nu_{\rm cut}=100~{\rm eV}$,
because higher-energy photons hardly ionize gas\footnote{
\cite{takeo_2019} do not set the cutoff, namely $\nu_{\rm cut}\rightarrow \infty$.
Therefore, $N_{\rm ion}$ is overestimated and the BH mass dependence is modest.
In fact, the numerical result obtained in \cite{takeo_2019} is nicely explained by the evaluation of 
$\dot{N}_{\rm ion}$ in Eq. (\ref{ndot_disc}).}.
Using the expression and setting $\dot{m}_{\rm BH}=1$, the ratio of $R_{\rm ion}/R_{\rm B}$ before the transition is evaluated as
\begin{align}
\frac{R_{\rm ion}}{R_{\rm B}} \propto \dot{N}_{\rm ion}^{1/3} n_{\infty}^{-2/3} M_{\rm BH}^{-1}
 \propto M_{\rm BH}^{(\alpha-3)/3} n_{\infty}^{-2/3},
\label{eq:RionRB}
\end{align}
where $\dot{N}_{\rm ion} \propto M_{\rm BH}^\alpha$ ($\alpha=1.3$ in Eq. \ref{ndot_disc} and 
$\alpha=1$ for power-law spectra).
By equating $R_{\rm ion}\simeq R_{\rm B}$, we obtain the critical density for the transition as 
\begin{align}
n_{\rm crit} \propto M_{\rm BH}^{-1+(\alpha-1)/2}.
\label{eq:RionRB}
\end{align}
Therefore, the critical condition is given by $M_{\rm BH}\times n_{\rm crit}  \propto M_{\rm BH}^{0.15}$ 
(blue curve in Fig. \ref{survey}) for disc spectra at $1< M_{\rm BH}/\msun < 10^6$ 
($M_{\rm BH}\times n_{\rm crit}  = {\rm const.}$ for power-law spectra; see green and black lines in Fig. \ref{survey}).
The lower critical density reflects that the radiation spectrum for lower BH mass is so hard 
that a significant fraction of photons do not contribute to ionization.

With mechanical feedback, the estimate of $\dot{N}_{\rm ion}$ is modified as follows.
As the outflow opening angle is wider and mass loading factor is higher, the BH accretion rate and radiation luminosity decrease.
We approximate the ionizing photon number flux as $\dot{N}_{\rm ion}\propto \dot{m}_{\rm BH}^{3/4}\propto [\Delta \Omega_{\rm in}
/(1+\beta_{\rm out})]^{3/4}$ (see discussion \S\ref{sec:out_para}).
Therefore, the size of ionized region is evaluated as
\begin{align}
R_{\rm ion} \propto \left[ \frac{\cos{\theta_{\rm out}}}{ \left( 1+\beta_{\rm out} \right) }\right]^{1/4} n_{\infty}^{-2/3},
\label{eq:RionRB}
\end{align}
where the part of $[...]$ is a correction factor from the mechanical feedback effect.
By equating $R_{\rm ion}\simeq R_{\rm B}$, we obtain the critical density with mechanical feedback as
\begin{equation}
n_{\rm crit,mf} \simeq \left[ \frac{\cos{\theta_{\rm out}}}{ \left( 1+\beta_{\rm out} \right)} \right]^{3/8} n_{\rm crit}
\equiv \mathcal{F}n_{\rm crit}.
\label{n_crit}
\end{equation}
For $\theta_{\rm out} = 80^{\circ}$ and $\beta_{\rm out}=1$, the correction factor is $\mathcal{F}\simeq 0.4$,
and the critical density is given as $n_{\rm crit,mf} \simeq 1.2 \times 10^4~\cc$ ($2.5 \times 10^7~\cc$) for 
$M_{\rm BH}=10^5~\msun$ ($10~\msun$), 
which nicely agrees to the simulation result (red curve in Fig. \ref{survey}).

We note that in the analytical argument above, we implicitly assume $\dot{M}_{\rm BH}\propto \Delta \Omega_{\rm in}$.
In fact, the simulation result indicates a modest $\Delta \Omega_{\rm in}$-dependence.
This is because gas in the equatorial region is compressed owing to expansion of the outflow region in
the tangential direction and the density of inflowing gas through the equator increases.
As the level of compression increases with the outflow opening angle, the dependence of $\dot{M}_{\rm BH}$ on 
$\Delta \Omega_{\rm in}$ becomes weaker than the linear relation.
However, gas compression makes radiative recombination more efficient simultaneously,
and promotes collapse of the ionized region.
As a result, our estimation assuming Eq. (\ref{n_crit}) nicely agrees with our numerical results,
although the compressional degree at the different radii is not modelled analytically.

%%%
\subsection{Typical outflow parameters}
\label{sec:outpara}

In our study, we investigate the impact of mechanical feedback on BH accretion, 
assuming the mass-loading factor and velocity at the injection scale.
We briefly summarize the outflow properties expected from recent numerical studies and 
discuss the uncertainties of the outflow model parameters.

As discussed in \S\ref{sec:tran}, the BH accretion rate is limited to $\dot{m}_{\rm BH}\lesssim 1$ before the transition.
In the sub-Eddington accretion phase, strong mass loss would be driven from the nuclear 
accretion disc scale via several different mechanisms.
A widely accepted idea is that the large scale magnetic field with poloidal topology plays a key role 
in acceleration and collimation of outflows and/or jets \citep[e.g.,][]{Blandford_1977,
Blandford_Payne_1982}.
Depending on the configuration and strength of magnetic field, there is a wide range of predictions
for the properties of outflows \citep[e.g.,][and references therein]{Jafari_2019}.
With a semi-analytical model of a sub-Eddington disc, \cite{Li_2019} find that strong mass loss driven 
by large-scale poloidal magnetic fields accumulated near the BH with accretion injects matter into 
outflows with a mass loading factor of $\beta_{\rm out} \approx 0.3-2$ (a higher $\beta_{\rm out}$
with stronger magnetic field).
The outflow velocity reaches $\simeq 0.1-0.2~c$ in the inner region and decreases outward.
In fact, a significant mass fraction of the outflow ($\gtrsim 50-70~\%$) has a relatively low velocity 
of $<0.01~c$, which is comparable to that in our fiducial value of $v_{\rm out}=1,000~{\rm km~s^{-1}}$.

Even in the sub-Eddington regime, radiation emitted from the nuclear disc can contribute to 
the acceleration of outflows by the line force due to bound-free absorption.
The line force increases by several orders of magnitude above the continuum radiation force exerted through electron scattering alone,
leading to high-speed outflows at $\simeq 0.05-0.1~c$ \citep{Proga_2000, Proga_Kallman_2004}.
A recent numerical study by \cite{Nomura_2018} found that the mass loading factor is as high as
$\beta_{\rm out}\lesssim 1.2$ when the mass inflow rate from larger radii is 
$0.3 \lesssim \dot{m}_{\rm in} \lesssim 1.5$.
In addition, X-ray emission from the inner disc and/or corona heats the surface of the disc 
at larger radii up to the Compton temperature \citep[e.g.,][]{Begelman_1983,Done_2018}.
The heated gas becomes unbound at the outer part of the disc and produces thermal-driven outflows
with a mass loading factor of $\beta_{\rm out} \sim O(1)$ at a terminal velocity of 
$v_{\rm out} \simeq 300-1,000~{\rm km~s^{-1}}$, which is the order of the sound speed 
of X-ray irradiated gas.

In the hyper-Eddington accretion phase after the transition, radiation force through 
electron scattering in optically-thick medium produces strong mass outflows from the accretion disc.
Numerical simulation studies suggested that the mass loading factor in the radiation-driven outflow
is as high as $\beta_{\rm out} \sim 0.05-0.3$ for $\dot{m}_{\rm BH} \sim 5-20$
\citep[][]{ohsuga+2009,Ohsuga_mineshige_2014,Jiang_2019}.
As the accretion rate rises, the value of $\beta_{\rm out}$ gradually increases; 
namely $\beta_{\rm out} \simeq 6-7$ for $\dot{m}_{\rm BH} \approx 10^2 $ \citep[][]{kawashima+2009}.
The outflow velocity is accelerated up to $\sim 0.2-0.5~c$ for $10 \lesssim \dot{m}_{\rm BH} \lesssim 10^2$
\citep{ohsuga+2009, kawashima+2009}.

In summary, the mass loading factor obtained from numerical and semi-analytical studies is 
$\beta_{\rm out} \sim 1$ and $\beta_{\rm out}\lesssim 10$ before and after the transition, respectively.
Depending on the outflow launching mechanisms and accretion rates,
the outflow velocity would be accelerated to several percent of light speed, 
which is commonly observed in luminous AGNs \citep{Tombesi_2010,Tombesi_2011}.
Since the typical velocity of outflows tends to higher than our fiducial value by one order of magnitude,
our result seems conservative for determining the criteria required for the hyper-Eddington transition
(see in \S\ref{sec:out_para}).

\subsection{Impact of BH feedback on the host galaxy evolution}

%% Fig. 11 %%
\begin{figure}
\begin{center}
\includegraphics[width=8.5cm]{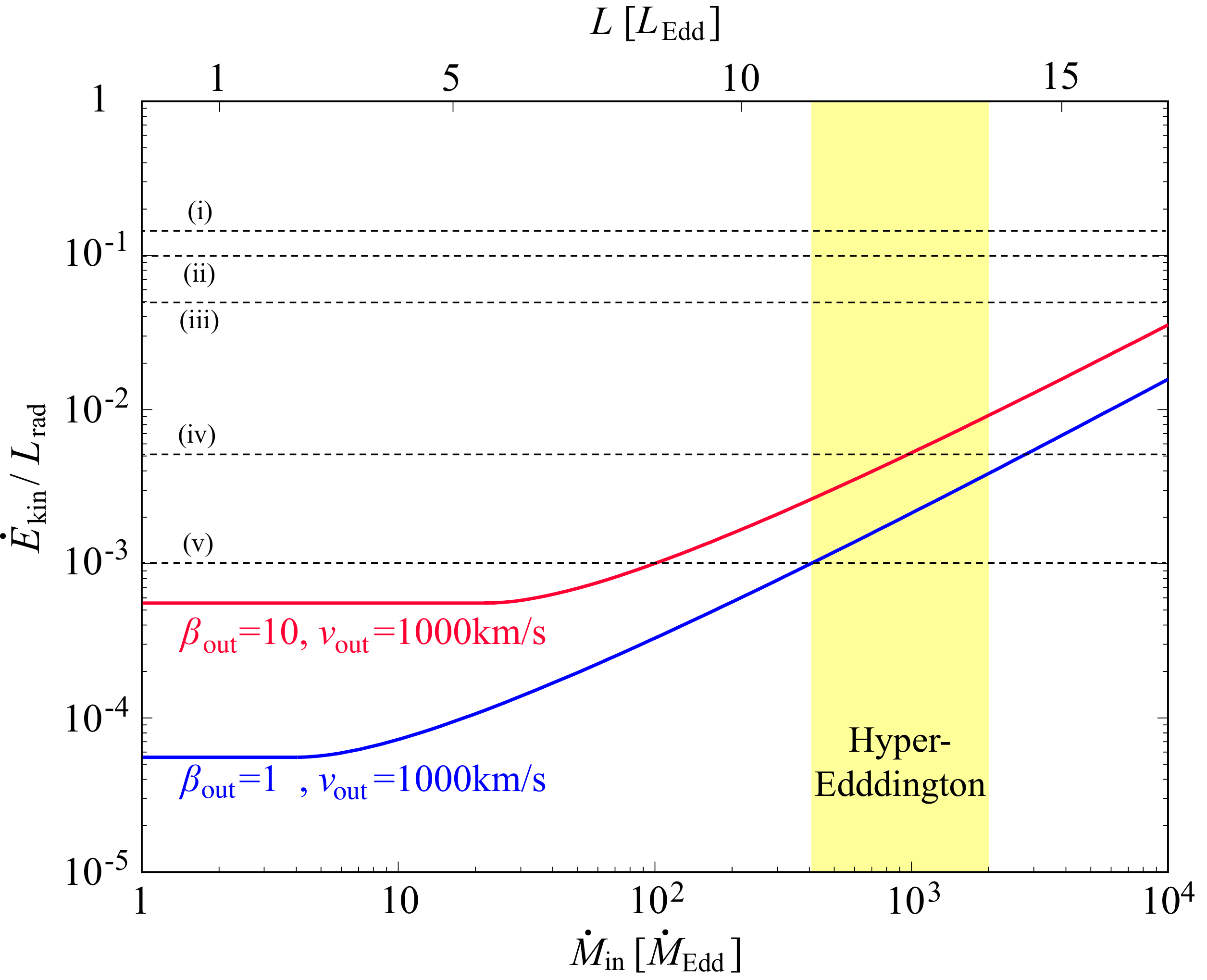}
\end{center}
\caption{
Kinetic coupling efficiency ($\dot{E}_{\rm out}/L_{\rm rad}$) in our feedback model for hyper-Eddington accretion
as a function of mass inflow rate $\dot{M}_{\rm in}$;
Model A45M5 ($\beta_{\rm out} =1$; blue) and A45M5-B10 ($\beta_{\rm out} =10$; red).
The upper horizontal axis shows the radiative luminosity (in units of $L_{\rm Edd}$) for $\beta_{\rm out} =1$. 
The yellow shaded region presents the mass inflow rates for hyper-Eddington accretion phases accompanied by powerful outflows.
Theoretical values adopted in cosmological simulations as sub-grid parameters are shown by horizontal lines: 
from the top to the bottom (i) \citet{Dubois+2014,Schaye+2015}, (ii)  \citet{Weinberger+2017},
(iii) \citet{DiMatteo+2005}, (iv) \citet{HopkinsElvis2010}, and (v) \citet{Costa+2018}.
}
\label{fig:kinrad}
\end{figure}

Rapid assembly of massive black holes via gas accretion is a key ingredient to understand 
the formation of high-$z$ SMBHs, which would require super/hyper-Eddington accretion but 
with a low duty cycle.
In a cosmological context, the rapid accretion mode would invoke the rapid collapse of chemically pristine primordial gas 
in so-called ``atomic cooling halos'' with virial temperature of $T_{\rm vir}\sim 10^4~\K$ \citep{VR_2005,inayoshi+16,Pezzulli_2016},
where a dense core region is developed at the nuclei of protogalaxies that do not experience prior star formation 
\citep{Wise+2008, Rega+2014}.
In recent decades, both semi-analytical models and cosmological simulations have been utilized to study the coevolution of SMBHs 
with their host galaxies, including various feedback processes associated with supernovae and AGN activity. 
Due to numerical limitations (e.g., spatial resolutions), however, a great number of studies treat feedback effects using sub-grid models 
instead of directly resolving physical processes within the Bondi scales.
In addition to feedback mechanisms, most cosmological simulations calculate the feeding rate onto unresolved BHs,
assuming $\dot{M}_{\rm BH}={\rm min}(\dot{M}_{\rm B},~ \dot{M}_{\rm Edd})$ without permitting super/hyper-Eddington accretion.
Therefore, the link between the model and physical processes of mass accretion and ejection from the nuclear region 
is still highly uncertain.

In studies of AGN feedback both observationally and theoretically, the kinetic coupling efficiency 
defined by the ratio of the kinetic luminosity of outflows to the AGN radiative luminosity, 
$\dot{E}_{\rm out}/L_{\rm rad}$, is often used to characterize the AGN outflow properties 
\citep[][reference therein]{Harrison+2018}.
In Fig. \ref{fig:kinrad}, we show the kinetic coupling efficiency calculated with the feedback model discussed in \S\ref{sec:2}
for two different outflow parameters (solid curves).
In the limit of high accretion rates ($\dot{m}_{\rm BH}\gg 1$), the value can be approximated as 
\begin{equation}
\frac{\dot{E}_{\rm out}}{L_{\rm rad}} 
\simeq \frac{5}{2}\beta_{\rm out}~ \frac{\dot{m}_{\rm BH}}{\ln \left(\dot{m}_{\rm BH}/2\right)}
\left(\frac{v_{\rm out}}{c}\right)^2,
\label{kinrad}
\end{equation}
where $\dot{m}_{\rm BH}=\dot{m}_{\rm in}/(1+\beta_{\rm out})$.
This relation weakly depends on the mass loading factor but strongly depends on the outflow velocity.
Since the range of mass inflow rates obtained from Model A45M5 ($\beta_{\rm out} =1$; blue)
and A45M5-B10 ($\beta_{\rm out} =10$; red) is $400<\dot{m}_{\rm in}<2000$ in their hyper-Eddington stages, 
the kinetic coupling efficiency is $\sim 0.1-1\%$ of the radiative luminosity.
We also overlay representative values for the kinetic coupling efficiencies 
adopted in cosmological simulations as sub-grid parameters, ranging 
$\sim 0.1-10\%$ of the radiative luminosity: 
from the top to the bottom (i) \cite{Dubois+2014,Schaye+2015}, (ii)  \cite{Weinberger+2017},
(iii) \cite{DiMatteo+2005}, (iv) \cite{HopkinsElvis2010}, and (v) \cite{Costa+2018}.
Although the fiducial values in our model are located at the lower end,
the mechanical power associated with highly accreting massive BHs, $\dot{E}_{\rm out}/L_{\rm rad}\simeq 10^{-3}-10^{-2}$, 
would impact upon their host galaxies with a significant delay time, which might be marginally
longer than the typical AGN lifetime \citep[e.g.,][]{Costa+2018}.
Finally, we note that the expression of Eq. (\ref{kinrad}) can be applied {\it only when the transition criterion for hyper-Eddington accretion is satisfied,
i.e., a mass inflow rate exceeding $\simeq 300~\dot{M}_{\rm Edd}$ is found at the inner-most cells large-scale cosmological simulations}.

%%%%%%%%%%%
%   5. Discussion    %
%%%%%%%%%%%
\section{Summary}
\label{sec:5}

We investigate the properties of accretion flows onto BHs at the nuclei of protogalaxies
and study the impact of mechanical and radiative feedback on rapid growth of BHs,
performing two-dimensional RHD simulations that take into account mechanical feedback with 
a phenomenological outflow model.
In our fiducial case ($\theta_{\rm out}=45^{\circ}$, $\beta_{\rm out}=1$, and $v_{\rm out}=1,000~{\rm km~s^{-1}}$),
we find that the flow structure consists of two distinct parts before the transition;
the bipolar outflowing region heated up to $T\sim 10^{6-7}~\K$ due to strong shock
and the equatorial inflowing region where ionized gas is mildly heated to $T\sim 10^5~\K$ 
due to photoionization. 
Since the outflows inject momentum and energy into the surrounding medium and prevent mass accretion onto the BH,
radiative output from the accreting BH is reduced. 
When the BH is embedded in a dense gas core, attenuated ionizing radiation hardly affects the gas dynamics 
at the Bondi radius, from which intense inflows of neutral gas occur at rates substantially exceeding the Eddington limit 
without impeded by photoionization and heating.
After the transition to the hyper-Eddington accretion phase, strong bipolar outflows completely evacuate 
the surrounding gas in the polar region but mass inflows through the equatorial region maintain the BH accretion rate 
as high as $\sim 300-10^3~\dot{M}_{\rm Edd}$, which is reduced by one order of magnitude from those 
with radiative feedback alone.
We note the mass inflow rate and flow structure in the rapidly accreting stage barely depend on the mass loading factor into outflows,
while more spherical radiation field makes the mass inflow highly episodic (the time-averaged accretion rate does not change significantly).

In order to derive the necessary conditions required for the transition, we conduct a comprehensive survey 
on the model parameter dependence, varying the outflow opening angle, mass-loading factor, 
outflow velocity, and density of gas surrounding the BH.
We find that the critical density above which the transition occurs is reduced by a factor of $\sim 3$,
nearly independently of BH mass, when mechanical feedback is considered.
We also conclude that contrary to naive expectation, the existence of stronger outflow (i.e., a wider opening angle, 
higher mass-loading factor, and higher outflow velocity) leads to the transition to rapid accretion phases more efficiently.
This is because the suppression of BH accretion owing to outflows reduces the radiative output from the nuclear BH, 
making a smaller ionized region surrounding the BH that collapse in a shorter dynamical timescale.

Finally, we discuss the application of our result as a sub-grid model for cosmological simulations that 
do not directly resolve the nuclear region scale.
Rapidly growing BHs inject mechanical power with $\sim 0.1-1\%$ of the radiative luminosity 
(see Eq. \ref{kinrad}) and would impact on their host galaxies.

\section*{Acknowledgements}

We would like to thank Ken Ohsuga and Hiroyuki R. Takahashi for providing their numerical code
and discussions. We also thank Kazuyuki Sugimura and Ryota Tomaru for fruitful discussion.
This work is partially supported by the National Science Foundation of China (11721303, 11991052, 11950410493), 
the National Key R\&D Program of China (2016YFA0400702), and by JSPS Grant-in-Aid for Scientific Research (C) (17K0583 SM).
Numerical computations were carried out on Cray XC50 at Center for Computational Astrophysics, 
National Astronomical Observatory of Japan, and High-performance Computing Platform of Peking University.

\bibliography{ref.bib}

%\appendix

% Don't change these lines
\bsp	% typesetting comment
\label{lastpage}
\end{document}